\documentclass[sigconf]{acmart}

\usepackage{graphicx}
\usepackage{amsmath}
\usepackage{array}
\usepackage{float}
\usepackage{booktabs}
\usepackage{multirow}
\usepackage{enumitem}
\usepackage{colortbl}
\usepackage{xcolor}
\usepackage{tcolorbox}
\usepackage{arydshln}
\usepackage{dashrule}

\title{Open FinLLM Leaderboard: Towards Financial AI Readiness}

\setcopyright{acmlicensed}

\copyrightyear{2024}
\acmYear{2024}
\setcopyright{acmlicensed}
\acmConference[ICAIF 24]{International Workshop on Multimodal Financial Foundation Models (MFFMs) at 5th ACM International Conference on AI in Finance}{Nov. 15}{NY, USA}
\acmBooktitle{International Workshop on Multimodal Financial Foundation Models (MFFMs) at 5th ACM International Conference on AI in Finance (MFFM at ICAIF '24), November 14--17, 2024, Brooklyn, NY, USA}

\begin{document}

\begin{teaserfigure}
  \includegraphics[width=\textwidth]{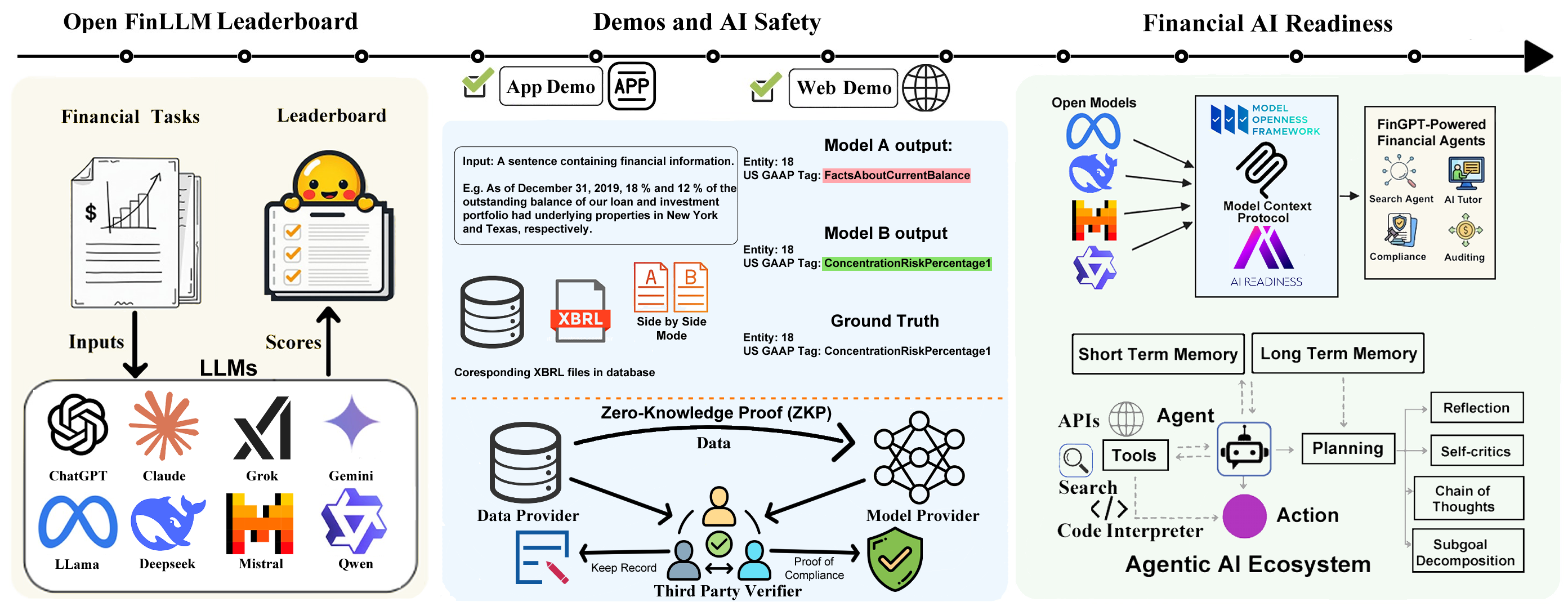}
  \caption{Overview of the open FinLLM leaderboard, from leaderboard, demos to financial AI readiness.}
  \label{fig:overview}
\end{teaserfigure}


\begin{abstract}

Financial large language models (FinLLMs) with multimodal capabilities are envisioned to revolutionize applications across business, finance, accounting, and auditing. However, real-world adoption requires robust benchmarks of FinLLMs' and FinAgents' performance. Maintaining an open leaderboard is crucial for encouraging innovative adoption and improving model effectiveness. In collaboration with Linux Foundation and Hugging Face, we create an open FinLLM leaderboard, which serves as an open platform for assessing and comparing AI models' performance on a wide spectrum of financial tasks. By demoncratizing access to advances of financial knowledge and intelligence, a chatbot or agent may enhance the analytical capabilities of the general public to a professional level within a few months of usage. This open leaderboard welcomes contributions from academia, open-source community, industry, and stakeholders. In particular, we encourage contributions of new datasets, tasks, and models for continual update. Through fostering a collaborative and open ecosystem, we seek to promote financial AI readiness.

\end{abstract}


\author{Shengyuan Colin Lin}
\authornote{Shengyuan Colin Lin worked as a RA at Columbia University during Fall 2024.}
\affiliation{\institution{Rensselaer Polytechnic Institute}
\country{USA}}
\email{lins10@rpi.edu}

\author{Felix Tian}
\authornote{Felix Tian contributed equally to this project.}
\affiliation{\institution{Rensselaer Polytechnic Institute}
\country{USA}}
\email{tianf2@rpi.edu}

\author{Keyi Wang}
\authornote{Keyi Wang contributed equally to this project. She worked as a RA at Columbia University.}
\affiliation{\institution{Columbia University}
\country{USA}}
\email{kw2914@columbia.edu} 

\author{Xingjian Zhao}
\affiliation{\institution{Rensselaer Polytechnic Institute}
\country{USA}}
\email{zhaox8@rpi.edu }

\author{Jimin Huang}
\affiliation{%
  \institution{FinAI}
  \country{Singapore}}
\email{jimin.huang@thefin.ai}

\author{Qianqian Xie}
\affiliation{%
  \institution{Yale University}
  \city{New Haven}
  \state{Connecticut}
  \country{USA}}
\email{qianqian.xie@yale.edu}

\author{Luca Borella}
\affiliation{%
  \institution{Linux Foundation}
  \city{New York}
  \country{USA}}
\email{luca.borella@finos.org}

\author{Matt White}
 \affiliation{%
    \institution{GM of AI, Linux Foundation; Executive Director, PyTorch Foundation; UC Berkeley}
   \city{Berkeley}
   \country{USA}
   \email{matt.white@linuxfoundation.org}
 }

\author{Christina Dan Wang}
\affiliation{%
  \institution{New York University Shanghai}
  \city{Shanghai}
  \country{China}}
\email{christina.wang@nyu.edu}

\author{Kairong Xiao}
\affiliation{%
  \institution{Business School, Columbia University}
  \city{New York}
  \state{NY}
  \country{USA}}
\email{kx2139@columbia.edu}

\author{Xiao-Yang Liu Yanglet}
\authornote{Corresponding author}
\affiliation{%
  \institution{Columbia University and Rensselaer Polytechnic Institute}
  \city{New York}
  \state{NY}
  \country{USA}}
\email{XL2427@columbia.edu}

\author{Li Deng}
\affiliation{
  \institution{University of Washington}
  \city{Seattle}
  \state{WA}
  \country{USA}
}
\email{deng629@gmail.com}

\maketitle
\section{Introduction}

Financial large language models (FinLLMs) and FinAgents with multimodal capabilities are rapidly advancing, poised to revolutionize a wide range of applications across business, finance, accounting, auditing, etc. FinLLM-powered agents streamline XBRL file analysis \cite{Han2024XBRLAgent}, automating financial filings and enhancing transparency across financial processes. FinGPT Search Agents \cite{Felix2024FinGPTAgent}, based on FinGPT \cite{Liu2023FinGPT, Liu2024FinGPTHPC, Yang2023FinGPT}, help decision-makers retrieve financial information with verified sources. FinLLMs show promise in social media analysis, such as understanding online interactions during the GameStop short squeeze event \cite{Lin2024GameStopEvent}. These models can also improve credit scoring, fraud detection, and regulatory compliance \cite{Nie2024SurveyLLMs,xie2023pixiu}, by improving risk assessments, fraud identification, and legal document processing in response to evolving regulatory demands \cite{wang2024report, keyi2024mffm, zhao2024revolutionizing}. More financial tasks can be found in \cite{Xie2024FinBen}.

Although FinLLMs and FinAgetns have great potentials, there are several major challenges to be addressed. One of the most impactful issues is ``hallucination" \cite{kang2023deficiency}, where models generate plausible but inaccurate information. Unreliable performance is an obstacle when it comes to financial decision-making as it would imply serious risks. LLMs also struggle to meet the complex demands of financial professionals who require high levels of precision, reliability, and advanced capabilities in quantitative reasoning and contextual accuracy. Privacy-preserving of training data, testing data, and model weights is another critical issue. Addressing these limitations requires advancements in multimodal FinLLMs, which are essential for financial tasks such as trading, risk management, and regulatory compliance \cite{ding2024large}.

In order to continually develop and fine-tune open multimodal LLMs, a well-accepted leaderboard with a standardized evaluation framework and continuous maintenance is essential. However, existing benchmarks, such as FinBen \cite{Xie2024FinBen} and FinanceBench \cite{Islam2023FinanceBench}, are static and lack the momentum to continuously adapt to and evaluate emerging FinLLMs and FinAgents, thereby limiting their utility for real-world applications and innovations.

In this paper, we present an open platform, \textit{FinLLM leaderboard}, that evaluates and compares FinLLMs and FinAgents across a wide spectrum of financial tasks. It is a collaborative project with Linux Foundation and Hugging Face. This leaderboard provides a transparent and standardized framework that ranks models based on their (multimodal) performance in areas such as financial reporting, sentiment analysis, and stock prediction. It also serves as an open platform for the community to evaluate, interact with, and compare FinLLMs and FinAgents in real-world scenarios. Beyond numeric scores, we showcase the integration with the FinGPT Search Agent \cite{Felix2024FinGPTAgent}, a promising use case of a personalized financial advisor. Users can explore, interact with, and compare models through demos. Additionally, we encourage contributions of models, datasets, and tasks to keep the leaderboard dynamic and responsive to the evolving needs of the financial industry. The leaderboard is continuously evolving, ensuring that it remains up-to-date with the latest FinLLMs and agents and adapts to more profesional-level financial tasks. We aim to foster an open collaborative ecosystem for long-term  maintenance by following the Model Openness Framework \cite{white2024model} \footnote{\url{https://isitopen.ai/}} and OpenMDW License\footnote{https://openmdw.ai/}. 

The educational documents of the open FinLLM leaderboard is available on this website\footnote{https://finllm-leaderboard.readthedocs.io}, while the associated codes are available on Hugging Face\footnote{\url{https://huggingface.co/spaces/finosfoundation/Open-Financial-LLM-Leaderboard}} and Github\footnote{\url{https://github.com/finos-labs/Open-Financial-LLMs-Leaderboard}}, respectively. The open-source FinGPT Search Agent \cite{Felix2024FinGPTAgent} is available on Github\footnote{\url{https://github.com/Open-Finance-Lab/FinGPT-Search-Agent}}.

The remainder of this paper is organized as follows. Section 2 reviews related works. Section 3 provides an overview of the open FinLLM leaderboard. Section 4 describes two demos, in particular, an app demo and a web demo of model evaluations, as well as several use scenarios. Section 5 discusses the hurdles to financial AI readiness.

\section{Related Works}

\subsection{Development of FinLLMs}

The proprietary BloombergGPT \cite{Wu2023BloombergGPT} was trained on a mixture of general data and financial data. It reported good performance on tasks such as sentiment analysis, question answering, and report summarization. 

The open-source FinGPT \cite{Liu2023FinGPT, Liu2024FinGPTHPC, Felix2024FinGPTAgent, Yang2023FinGPT} aims to democratize access to FinLLMs and agents by offering an automatic data curation pipeline and releasing fine-tuned model weights. These FinLLMs emphasize transparency and community-driven development, offering open alternatives to proprietary models. Multimodal FinLLMs and agents, capable of integrating text, tables, and time-series data, have shown notable improvements in stock movement forecasting and risk management tasks \cite{Xie2024OpenFinLLMs}.

\begin{figure*}[t]
\centering
\includegraphics[width=\textwidth]{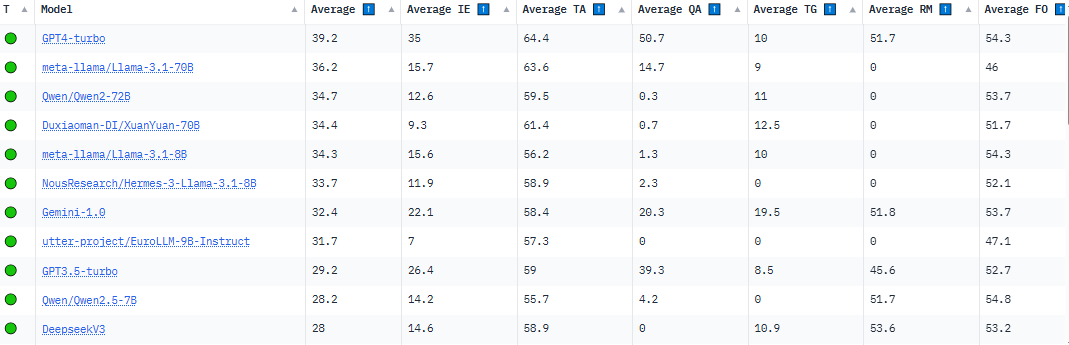}
\caption{A screenshot of the open FinLLM leaderborad. The top $11$ models are ranked across $7$ financial tasks.}
\label{fig:result_table}
\vspace{-0.1in}
\end{figure*}

\subsection{Benchmarking FinLLMs}
Various benchmarks have been developed to evaluate FinLLMs performance on financial tasks. One of the most comprehensive benchmarks is FinBen \cite{Xie2024FinBen}. The FinBen spans 24 tasks across 46 datasets, providing a robust framework for evaluating models like GPT-4 and Gemini. Tasks within FinBen cover information extraction, question answering, stock trading, and so on. Another notable benchmark is FinanceBench \cite{Islam2023FinanceBench} that focused on financial question answering and released 150 questions. Moreover, computational benchmarks such as FinGPT-HPC \cite{Liu2024FinGPTHPC} aimed to improve the efficiency of pretraining and fine-tuning LLMs for financial applications, emphasizing GPU memory optimization and training scalability in high-performance computing environments. Although tasks covering can be considerably comprehensive at the moment, benchmarks like FinanceBench are static in nature and lack interactive capabilities. Therefore, less likely to be well accepted and continuously contribute to the community.

Our open FinLLM leaderboard is not designed to compete with these benchmarks. We are moving one step further to complement them by offering an open platform for continuous contributions. We encourage collaboration with the open source comminuties and benchmark developers and integrate FinBen \cite{Xie2024FinBen} and FinanceBench \cite{Islam2023FinanceBench}, creating an evaluation ecosystem.

\section{Open FinLLM Leaderboard}

First, we provide an overview of the open FinLLM leaderboard. Then, we describe the financial tasks and testing pipeline.
 
\subsection{Overview}

Our work extends beyond setting up a leaderboard demo to provide benchmark results. In Fig. ~\ref{fig:overview} shows three stages of this leaderboard. We believe that the leaderboard serves as a crucial step towards financial AI readiness.

The open FinLLM leaderboard is collecting research lab-contributed financial evaluation tasks tested with pretrained and finetuned models, such as ChatGPT, LLaMA3, and Gemini.

We have a demo hosted on HuggingFace and a FinGPT Search Agent application \cite{Felix2024FinGPTAgent}, which serves as an interactive layer where users can access the leaderboard, view model performances, and directly compare results in a side-by-side view. 

We aim to identify financial AI readiness by delivering valuable insights for model deployment in real-world applications. It may help public users to identify models that outcompete others in certain financial tasks, giving a reliable standard for users choosing appropriate FinLLMs to address their specific needs. As a result, we can have a robust, AI-ready framework that empowers financial professionals with the right tools to enhance decision-making and operational efficiency.

\begin{table*}[htbp]
\centering
\caption{Financial tasks in the open FinLLM leaderboard.}
\label{tab:task_categories}
\begin{tabular}{|l|l|}
\hline
\textbf{Category}            & \textbf{Tasks}                                                                        \\ \hline
\textbf{Information Extraction (IE)}  & Named Entity Recognition (NER), Relation Extraction, Causal Classification.         \\ \hline
\textbf{Textual Analysis (TA)}        & Sentiment Analysis, Hawkish-Dovish Classification, Argument Unit Classification.               \\ \hline
\textbf{Question Answering (QA)}      & Answering financial questions from datasets like FinQA and TATQA.                              \\ \hline
\textbf{Text Generation (TG)}         & Summarization of financial texts (e.g., reports, filings).                          \\ \hline
\textbf{Risk Management (RM)}         & Credit Scoring, Fraud Detection, evaluating financial risks.                        \\ \hline
\textbf{Forecasting (FO)}             & Stock Movement Prediction based on financial news and social media.                            \\ \hline
\textbf{Decision-Making (DM)}         & Simulating decision-making tasks, e.g., M\&A transactions, trading tasks.                       \\ \hline
\end{tabular}
\end{table*}

\subsection{Financial Tasks with Multimodal Data}

We compare FinLLMs across multiple task categories including information extraction (IE), textual analysis (TA), question answering (QA), text generation (TG), risk management (RM), forecasting (FO) and Decision-Making (DM). It provides a comprehensive evaluation on tasks with multimodal data, such as text, tables, numerical data, and structured formats like XBRL. A screenshot of the scores is shown in Fig.~\ref{fig:result_table}.

\begin{figure}[t]
\centering
\includegraphics[width=0.5\textwidth]{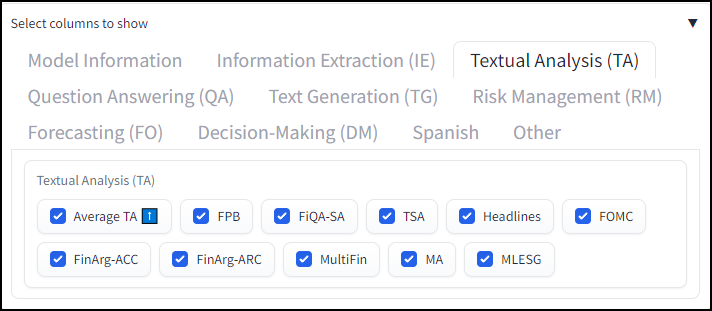}
\caption{Example of task selection, allowing users to browse tasks under different financial categories.}
\label{fig:task_selection}
\vspace{-0.1in}
\end{figure}

The open FinLLM leaderboard enables users to view and compare top-performing models. As shown in Fig.~\ref{fig:task_selection}, users have the flexibility to filter and reorder the displayed information by selecting specific categories and tasks. The current $42$ financial datasets are organized into seven categories, as given in Table \ref{tab:task_categories}.
\begin{itemize}[leftmargin=*]
    \item \textbf{Information extraction} (IE): Involves transforming unstructured financial data into structured formats, such as extracting entities or identifying relationships in financial agreements. It is essential for automating information retrieval of financial documents, such as SEC 10K filings.
    
    \item \textbf{Textual analysis} (TA): Involves evaluating how well LLMs quantify sentiment, classify financial news, or identify argumentative structures, which are crucial for market sentiment analysis.
    
    \item \textbf{Question answering} (QA): Involves evaluating models on their ability to answer complex financial queries, particularly involving numerical reasoning or document comprehension.
    
    \item \textbf{Text generation} (TG): Involves summarization tasks focus on generating coherent, concise representations of financial documents. It is important for creating reports or summaries from lengthy financial articles.
    
    \item \textbf{Risk Management} (RM): Involves predicting credit risk or detecting fraudulent behavior, which is critical in assessing the likelihood of default or fraud.
    
    \item \textbf{Forecasting} (FO): Involves tasks such as stock movement prediction which challenge models to anticipate market trends based on financial data. Effective forecasting has a direct impact on financial decision-making.
    
    \item \textbf{Decision-Making} (DM): Involves tasks that pertain to trading decisions or M\&A deal completeness and the ability to simulate and support decision-making in financial environments.
\end{itemize}

Users can easily explore these categories and customize the displayed results based on their interests, ensuring relevant and practical insights for financial analysis, as shown in Fig.~\ref{fig:task_selection}.

\subsection{Testing Pipeline}

\begin{figure*}[t]
\centering
\includegraphics[width=\textwidth]{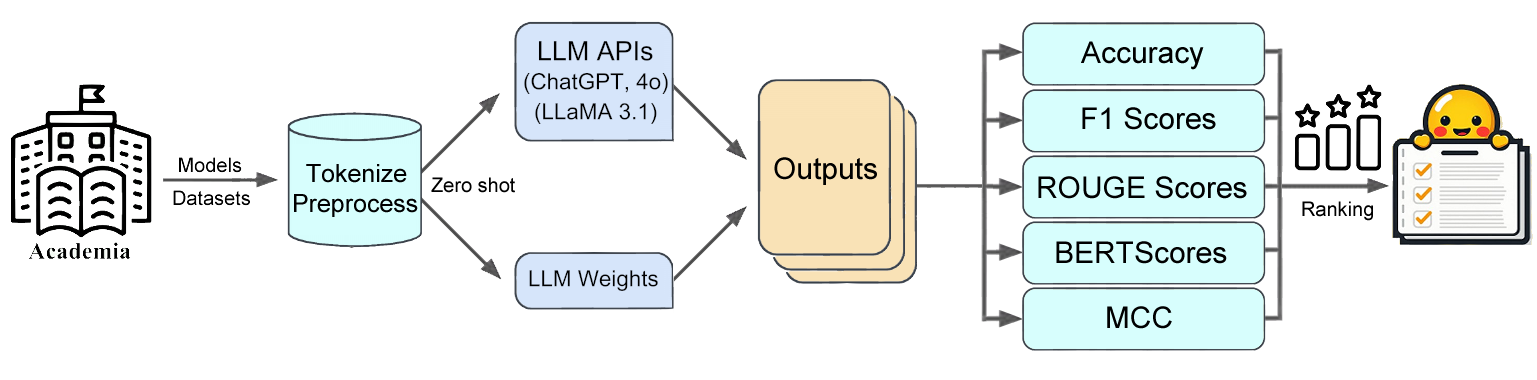}
\caption{Testing pipeline currently used in the FinLLM leaderboard.}
\label{fig:evaluation_pipeline}
\vspace{-0.1in}
\end{figure*}

Fig. \ref{fig:evaluation_pipeline} is an overview of our testing pipeline. We apply a zero-shot evaluation setting in testing expert-validated datasets, evaluating models in multimodal settings across various financial tasks. Models are compared fairly based on their ability to handle unseen tasks in finance. The evaluation code is available On Github \footnote{\url{https://github.com/finos-labs/Open-Financial-LLMs-Leaderboard}}.

The following popular models were evaluated in the current pipeline: GPT-4 (standard version) \cite{achiam2023gpt}, LLaMA 3.1 (both 8B and 70B versions) \cite{dubey2024llama}, Gemini \cite{anil2023gemini}, Qwen2 (72B and 7B-Instruct versions) \cite{yang2024qwen2}, Xuanyuan-70B \cite{zhang2023xuanyuan}.

\textbf{Testing pipeline}:
\begin{itemize}[leftmargin=*]
    \item \textbf{Model downloading}: Models are either downloaded from Huggingface or accessed via APIs. 
    \item \textbf{Preprocessing and tokenization}: Financial documents are tokenized to meet each model’s format and token limit requirements.
    \item \textbf{Zero-shot evaluation}: Models are evaluated without prior fine-tuning on the task-specific datasets, focusing on their generalization ability in financial contexts.
    \item \textbf{Expert-validated datasets}: The datasets are selected and validated by financial professionals, ensuring their relevance to real-world financial applications.
    \item \textbf{Metric calculation and normalization}: Each model's performance is measured across different metrics based on task requirements (Accuracy, F1 Score, ROUGE, BERTScore, MCC, etc.). For  fair comparison, all scores are normalized into the range \([0, 100]\) using min-max scaling:
    \begin{equation}
    \overline{S} = \frac{S - \text{min}}{\text{max} - \text{min}} \times 100,
    \end{equation}
    where \(S\) is the raw score, and \([\min, \max]\) is the original range. For example, a score of $0$ in \([-3, 3]\) normalizes to $50$, while $0.5$ in \([0, 1]\) also normalizes to $50$.
    \item \textbf{Ranking}: After normalization, models are ranked based on their scores across different tasks, providing an aggregate performance metric for comparison.
\end{itemize}

\textbf{Evaluation metrics}. The models' performance is measured
using the following metrics:
\begin{itemize}[leftmargin=*]
    \item \textbf{Accuracy}: Used in the credit scoring and Hawkish-Dovish classification tasks.
    \item \textbf{F1 score}: Used in the sentiment analysis, named entity recognition (NER), and relation extraction tasks.
    \item \textbf{ROUGE score}: Used
     in the summarization tasks, measuring the quality of generated text by comparing it with reference summaries.
    \item \textbf{BERTScore}: Measures the similarity between the generated and reference summaries at a more granular, contextual level.
    \item \textbf{Matthews correlation coefficient (MCC)}: Used in the binary classification tasks, such as fraud detection and credit scoring.
\end{itemize}

\begin{figure}[t]
\centering
\includegraphics[width=0.5\textwidth]{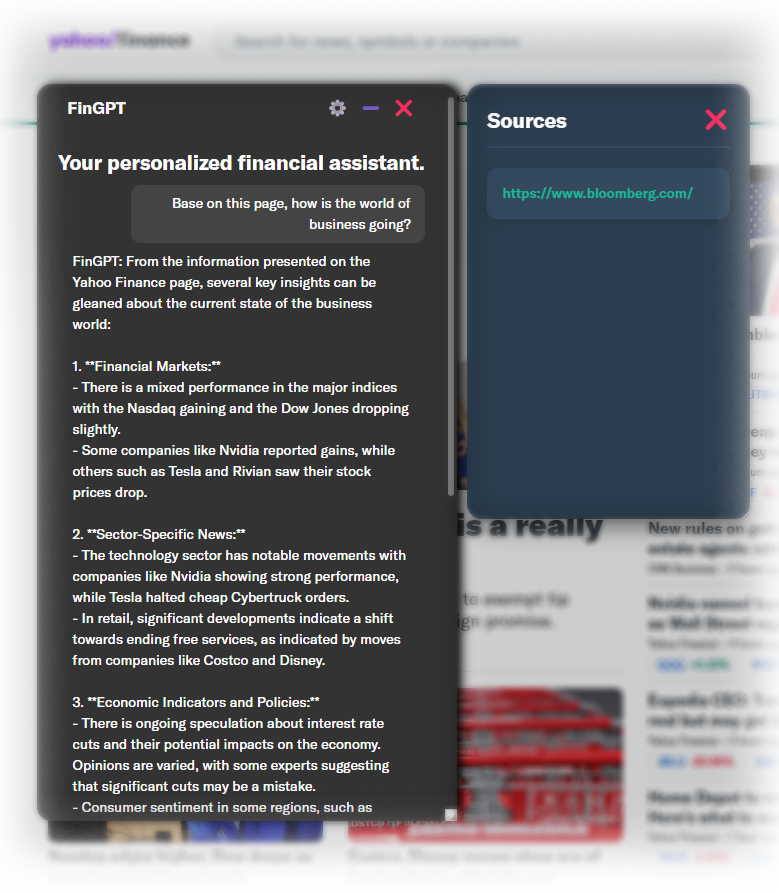}
\caption{Demo of FinGPT Search Agent \cite{Felix2024FinGPTAgent}: users could check information sources.}
\label{fig:search_agent}
\vspace{-0.1in}
\end{figure}

\subsection{Structure}
\begin{figure}[t]
\centering
\includegraphics[width=0.5\textwidth]{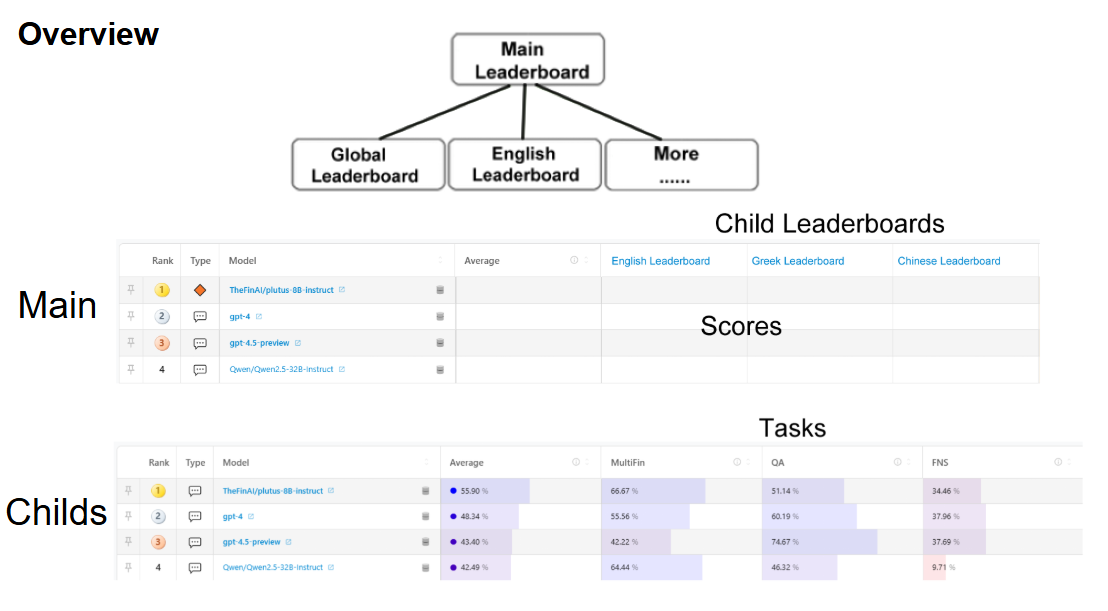}
\caption{Global performance overview showing weighted average scores across all scenarios}
\label{fig:tree_structure}
\vspace{-0.1in}
\end{figure}

Fig. \ref{fig:tree_structure} shows our hierarchical structure leaderboard system:

\begin{itemize}[leftmargin=*]
    \item \textbf{Main Leaderboard}: Aggregates performance across all application scenarios
    \item \textbf{Application-Specific Leaderboards}: Group tasks by real-world use cases
\end{itemize}

Each child leaderboard represents a core business application: 
\begin{itemize}[leftmargin=*]
    \item \textbf{Search Agent}: Enterprise search and information retrieval
    \item \textbf{AI Tutor}: Educational assistance and content generation
    \item \textbf{Compliance}: Regulatory monitoring and policy adherence
    \item \textbf{Auditing}: Financial and operational verification
\end{itemize}

This structure clearly maps child leaderboard to business functions, enabling the industries to find the scenarios they need and access to relevant benchmarks easily.

\section{Demos and Use Scenarios}

We present two types of demos for users to explore, utilize, and compare FinLLM models in various use scenarios.

\subsection{App Demo: Search Agent}

The FinGPT Search Agent \cite{Felix2024FinGPTAgent} demonstrates advanced capabilities in financial data retrieval and analysis by leveraging Retrieval-Augmented Generation (RAG). RAG integrates real-time information retrieval into the generative process, enabling the model to incorporate relevant, up-to-date data from sources such as Yahoo Finance, Bloomberg, or local documents like PDFs and Excel sheets \cite{zhao2024retrieval}. Fig.~\ref{fig:search_agent} illustrates the FinGPT Search Agent's capability of performing RAG and online search, ensuring that its responses are enriched with up-to-date and most relevant financial information.

\begin{figure*}[h]
    \centering
    \includegraphics[width=\textwidth]{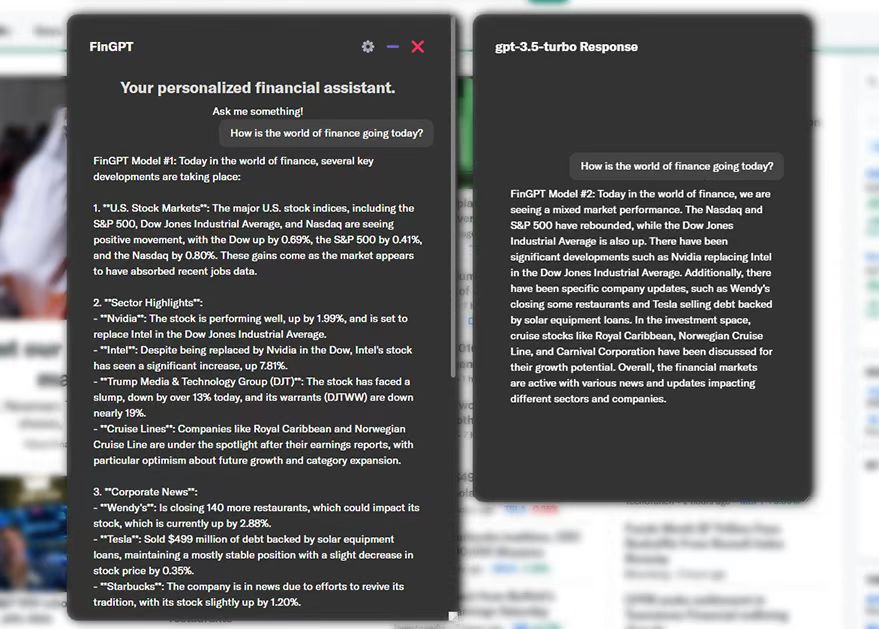}
    \caption{The side-by-side comparison mode of FinGPT Search Agent \cite{Felix2024FinGPTAgent}. Both Model \#1 and  Model \#2 respond to the prompt "How is the world of finance going today?".}
    \label{fig:side_by_side}
\end{figure*}

Fig.~\ref{fig:side_by_side} shows a side-by-side comparison mode of two selected models from the open FinLLM leaderboard. Both models utilize the same set of verifiable sources, enabling users to directly compare their results. This integration not only allows for a comparative analysis using performance metrics from the open FinLLM leaderboard but also facilitates source verification, ensuring that the generated responses are grounded in reliable, up-to-date information. The demo provides a hands-on experience, allowing users to assess the strengths and weaknesses of different models in real-world financial contexts. 

\textbf{Side-by-side comparison mode}. 
 Fig.~\ref{fig:side_by_side} showcases FinGPT search agent's side-by-side comparison feature \cite{Felix2024FinGPTAgent}, displaying responses from Model 1 and Model 2 to the prompt, "How is the world of finance going today?". This format enables users to qualitatively assess differences in response style, depth, and relevance, providing insights beyond traditional evaluation scores. While standard metrics like accuracy offer limited insights, our open-box evaluation allows users to explore model outputs directly. By presenting responses side-by-side, FinGPT’s leaderboard goes “beyond the scores,” helping users evaluate practical aspects such as clarity, factual accuracy, and relevance in real-world contexts. This approach opens the black box of model evaluation, aligning metrics with real-world utility.


Fig.~\ref{fig:OFLLM_Benchmark_Diagram} demonstrates the unique workflow of the Open Financial LLM Leaderboard. Beyond presenting task-specific scores, it includes a powerful side-by-side comparison feature. This functionality allows users to:
\begin{itemize}
    \item Select two models to compare.
    \item Provide a single test input to both models.
    \item Observe and evaluate the outputs generated by the models side-by-side.
\end{itemize}

\subsection{Web Demo}

The open FinLLM leaderboard demo, hosted on Huggingface with support from the Linux Foundation, provides users an intuitive way to interact with a variety of financial tasks and an easy-to-navigate environment for evaluating financial language models. Users can select tasks from multiple categories, such as Information Extraction, Risk Management, and Forecasting, and assess the performance of various models across these tasks. Fig.~\ref{fig:task_selection} provides a screenshot of the task selection interface.

Once a task is selected, users can view a comprehensive table summarizing model performance using relevant metrics. This setup enables direct comparisons across different models and tasks, helping users identify the most suitable models for their specific needs. Fig.~\ref{fig:result_table} illustrates an example of the result table, which presents the performance metrics for easy comparison and reference.

\subsection{Zero-Knowledge Proof}
In the open FinLLM leaderboard, we apply the Zero-Knowledge Proof (ZKP) technology. This feature aims to protect the privacy of the dataset and prevent fraudulent behaviors such as leaderboard hacking. With ZKP, we envision a system that can verify model performance without exposing sensitive input data, ensuring both integrity and security of the evaluation process.

Our Zero-Knowledge Proof (ZKP) implementation ensures evaluation integrity while protecting sensitive data:
\begin{itemize}[leftmargin=*]
    \item \textbf{Privacy-Preserving Verification}: Models can prove their performance without exposing training data
    \item \textbf{Anti-Gaming Protection}: Prevents leaderboard manipulation through cryptographic verification.
    \item \textbf{Data Confidentiality}: Financial institutions can contribute proprietary datasets without disclosure
    \item \textbf{Transparent Auditing}: All evaluations are cryptographically verifiable while maintaining privacy

\end{itemize}

\subsection{Use Scenarios}

The open FinLLM leaderboard would potentially benefit the following financial applications. Here, we provide exploration and analysis of real-world use scenarios, inspired by \cite{yang2023dawnlmmspreliminaryexplorations}. 

\subsubsection{Refining Questions for Legal Consultations}
FinLLMs can help users refine their inquiries before meeting with legal counsel, leading to significant time and cost savings. By generating focused and precise questions in advance, clients reduce consultation time, thereby enhancing the efficiency and cost-effectiveness of legal interactions. On the service provider side, lawyers benefit from streamlined case preparation, allowing for the rapid identification of pertinent cases and regulations, which ultimately enhances productivity. This approach can be generalized to other domains, including medical consultations, exam preparation, and the analysis of records. An illustrative example of this application is provided in Appendix Fig.~\ref{fig:lawyer-use-scenario-example}.

\subsubsection{General Public vs. LLM-Assisted Financial Document Analysis}
FinLLMs empower the general public to comprehend complex financial documents, such as earnings reports or regulatory filings, with a level of proficiency comparable to that of financial professionals. By simplifying, summarizing, and contextualizing information, these models enable users to make more informed decisions. Whether analyzing investment opportunities or assessing corporate financial health, FinLLMs help demystify intricate financial data, thereby bridging the knowledge gap between professionals and non-specialists. An illustrative example is provided in appendix Fig. ~\ref{fig:report_use_scenario_example}

\subsubsection{Simplifying Financial Analysis for Everyday Users}
Platforms such as Yahoo Finance, Bloomberg, The Wall Street Journal, Business Insider, MarketWatch, and CNBC frequently present vast and often overwhelming amounts of financial information. FinLLMs assist users by extracting key data points and insights, effectively addressing questions such as, 'What are the most critical elements on this page to consider if I am evaluating potential investments?' This functionality not only saves users time but also enhances their comprehension of complex financial content, thereby empowering more informed investment decisions. An illustrative example of this application is provided in Appendix Fig. ~\ref{fig:report_use_scenario_example}.



\section{Towards Financial AI Readiness}

Our goal would be to build an open community that pushes financial AI to be ready for real-world applications and to build a gateway between academia and industry. By translating complex research achievements into accessible and actionable insights, we foster the growth of the Agentic AI Ecosystem. The Open FinLLM Leaderboard is similar to established industry standards such as MCP and MOF. We set the benchmark for financial AI readiness, ensuring that innovations in financial language models are both practical and impactful. In this section, we discuss critical aspects and how this leaderboard and the surrounding community will contribute to FinLLweMs' readiness.

\subsection{Evaluating Financial Performance}

Financial AI readiness requires FinLLM models that not only perform technically well but also integrate with existing workflows, provide useful insights, and comply with industry regulations. The open FinLLM leaderboard assesses how models perform on financial tasks like risk management, sentiment analysis, and regulatory compliance. These evaluations help developers and practitioners select models for deployment and identify aspects to be improved.

By including a range of tasks that reflect real financial challenges, this leaderboard ensures models are capable of handling financial analytical problems, not just standard text tasks. This supports the potential integration of models into financial decision-making processes, such as investment analysis, compliance checks, and automated reporting.

\subsection{Revealing Model Limitations: Hallucinations and Interpretability}

Accuracy and reliability are crucial in terms of financial applications. However, A major challenge existed is the risk of "hallucinations," where models produce incorrect or misleading information. Affecting by AI generated misinformation, people tend to make flawed investment decisions or misunderstandings of regulations \cite{kang2023deficiency}.

To identify whether AI is hallucinating, we include tests focused on interpretability and reliability specifically for financial services. After we evaluate models on tasks where precision is critical, we can help financial institutions identify and reduce risks. In the future, stakeholders and financial institutions can deploy models fitting well on their demands by assessing between model evaluation results.

\subsection{Ethical Considerations and Transparency}
High ethical standards is also essential for AI to be a part of financial decision-making. In this case, the open FinLLM leaderboard encourages transparency, fairness, and ethical compliance in deploying financial AI by providing a specialized evaluation framework in financial sector \cite{Uzougbo2024}.

Our zero-shot evaluation method tests models on new tasks, simulating the real-world financial scenarios where models usually given tasks without observing from examples. This approach can reveal a model’s actual abilities and limitations, therefore building trust among financial institutions, regulators, and the public. The leaderboard offers transparent evaluations and can helps organizations make informed decisions and meeting regulatory and ethical standards when using AI. 

\subsection{Future Research}

\textbf{Multimodal capabilities} \cite{liu2024MFFM}:  For financial AI readiness, models need to handle different types of data, such as text, figures, tables, time-series data, and alternative data.  The leaderboard will keep evolving to meet new challenges, through community involvement and continuous updates.

\textbf{GPU optimization for inference}: To provide personalized financial advices \cite{Felix2024FinGPTAgent} (running locally on personal devices), it is important to reduce GPU memory consumption and response time. LoRA fine-tuning is practically useful for adapting a general purpose model to a personalized version, such as the ongoing project FinLoRA \cite{wang2024finlora}, while uantization into 8 bit or 4 bits is an effective technique that greatly reduce GPU memory consumption \cite{Liu2024FinGPTHPC}.  

\textbf{Zero-Knowledge Proofs (ZKPs)}: The privacy concern is critical in business and finance, while the inference process of our leaderboard is no exception. One the one hand, close-source models require the protection of model weights and training datasets (possible other artifacts). ZKPs provide a verifiable auditing process for newly released model weights. On the other hand, we expect to include evaluation results on proprietary testing datasets. ZKPs allow authors to publish such results along with the generated proof-file, without revealing information of the testing dataset.


\begin{acks}
Shengyuan Colin Lin, Keyi Wang, and Xiao-Yang Liu Yanglet acknowledge the support from Columbia's SIRS and STAR Program, The Tang Family Fund for Research Innovations in FinTech, Engineering, and Business Operations. Shengyuan Colin Lin, Felix Tian, Xingjian Zhao, and Xiao-Yang Liu Yanglet acknowledge the support from NSF IUCRC CRAFT center research grant (CRAFT Grant 22017) for this research. The opinions expressed in this publication do not necessarily represent the views of NSF IUCRC CRAFT. 
\end{acks}

\bibliographystyle{ACM-Reference-Format}
\bibliography{references}

\setcounter{section}{0}
\renewcommand\thesection{\Alph{section}}

\appendix 

\begin{table*}[htbp]
\centering
\setlength{\tabcolsep}{4pt}
\renewcommand{\arraystretch}{1}
\caption{Overview of financial tasks, evaluation metrics, and dataset sizes.}
\label{tab:combined_tasks}
\begin{tabular}{|p{1.5cm}|p{2.5cm}|p{4.5cm}|p{4.5cm}|p{1cm}|}
\hline
\textbf{Category} & \textbf{Dataset} & \textbf{Task} & \textbf{Evaluation Metric} & \textbf{Test Size} \\
\hline
\textbf{IE} & NER \cite{salinas-alvarado-etal-2015-domain}& Named Entity Recognition & Entity F1 & 980 \\
\textbf{IE} & FiNER-ORD \cite{shah2023finer} & Named Entity Recognition & Entity F1 & 1080 \\
\textbf{IE} & FinRED \cite{sharma2022finred} & Relation Extraction & F1, Entity F1 & 1068 \\
\textbf{IE} & SC \cite{mariko2020financial} & Causal Classification & F1, Entity F1 & 864 \\
\textbf{IE} & CD \cite{mariko2020financial} & Causal Detection & F1, Entity F1 & 226 \\
\textbf{IE} & FNXL \cite{sharma-etal-2023-financial} & Numeric Labeling & F1, EM Accuracy & 318 \\
\textbf{IE} & FSRL \cite{lamm2018textual} & Textual Analogy Parsing & F1, EM Accuracy & 97 \\
\hline
\textbf{Total IE} & & & & 4633 \\
\hline
\textbf{TA} & FPB \cite{malo2014good} & Sentiment Analysis & F1, Accuracy & 970 \\
\textbf{TA} & FiQA-SA \cite{10.1145/3184558.3192301} & Sentiment Analysis & F1 & 1173 \\
\textbf{TA} & TSA \cite{10.1007/978-3-319-69146-6_11} & Sentiment Analysis & F1, Accuracy & 561 \\
\textbf{TA} & Headlines \cite{10.1007/978-3-030-73103-8_41} & News Headline Classification & Avg F1 & 11412 \\
\textbf{TA} & FOMC \cite{shah-etal-2023-trillion} & Hawkish-Dovish Classification & F1, Accuracy & 496 \\
\textbf{TA} & FinArg-ACC \cite{sy2023fine} & Argument Unit Classification & F1, Accuracy & 969 \\
\textbf{TA} & FinArg-ARC \cite{sy2023fine} & Argument Relation Classification & F1, Accuracy & 690 \\
\textbf{TA} & MultiFin \cite{jorgensen-etal-2023-multifin} & Multi-Class Classification & F1, Accuracy & 546 \\
\textbf{TA} & MA \cite{yang2020generating} & Deal Completeness Classification & F1, Accuracy & 500 \\
\textbf{TA} & MLESG \cite{chen2023multi}& ESG Issue Identification & F1, Accuracy & 300 \\
\hline
\textbf{Total TA} & & & & 17617 \\
\hline
\textbf{QA} & FinQA \cite{chen2021finqa} & Question Answering & EM Accuracy & 1147 \\
\textbf{QA} & TATQA \cite{zhu2021tat} & Question Answering & F1, EM Accuracy & 1670 \\
\textbf{QA} & Regulations \cite{Xie2024FinBen} & Long-Form QA & ROUGE, BERTScore & 200 \\
\textbf{QA} & ConvFinQA \cite{chen2022convfinqa} & Multi-Turn QA & EM Accuracy & 1490 \\
\hline
\textbf{Total QA} & & & & 4507 \\
\hline
\textbf{TG} & ECTSum \cite{mukherjee2022ectsum} & Text Summarization & ROUGE, BERTScore, BARTScore & 495 \\
\textbf{TG} & EDTSum \cite{Xie2024FinBen} & Text Summarization & ROUGE, BERTScore, BARTScore & 2000 \\
\hline
\textbf{Total TG} & & & & 2495 \\
\hline
\textbf{FO} & BigData22 \cite{10020720} & Stock Movement Prediction & Accuracy, MCC & 1470 \\
\textbf{FO} & ACL18 \cite{xu-cohen-2018-stock} & Stock Movement Prediction & Accuracy, MCC & 27053 \\
\textbf{FO} & CIKM18 \cite{10.1145/3269206.3269290} & Stock Movement Prediction & Accuracy, MCC & 4967 \\
\hline
\textbf{Total FO} & & & & 33490 \\
\hline
\textbf{RM} & German \cite{hofmann1994statlog} & Credit Scoring & F1, MCC & 1000 \\
\textbf{RM} & Australian \cite{quinlan1987statlog} & Credit Scoring & F1, MCC & 690 \\
\textbf{RM} & LendingClub \cite{feng2023empowering} & Credit Scoring & F1, MCC & 13453 \\
\textbf{RM} & ccf \cite{feng2023empowering}& Fraud Detection & F1, MCC & 11392 \\
\textbf{RM} & ccfraud \cite{feng2023empowering}& Fraud Detection & F1, MCC & 10485 \\
\textbf{RM} & polish \cite{feng2023empowering}& Financial Distress Identification & F1, MCC & 8681 \\
\textbf{RM} & taiwan \cite{feng2023empowering}& Financial Distress Identification & F1, MCC & 6819 \\
\textbf{RM} & ProtoSeguro \cite{feng2023empowering} & Claim Analysis & F1, MCC & 11904 \\
\textbf{RM} & travelinsurance \cite{feng2023empowering}& Claim Analysis & F1, MCC & 12665 \\
\hline
\textbf{Total RM} & & & & 77089 \\
\hline
\textbf{DM} & FinTrade \cite{Xie2024FinBen}& Stock Trading & CR, SR, DV, AV, MD & 3384 \\
\hline
\end{tabular}
\end{table*}

\newpage

\section{Benchmark Datasets}

This section provides an overview of the current financial tasks in the open FinLLM leaderboard.  Table~\ref{tab:combined_tasks} provides an overview of the datasets, tasks, evaluation metrics, and test sizes, while Table~\ref{tab:task_sources} provides the corresponding sources. 


\subsection{Insights}

The categorization of datasets offers several important takeaways for the financial industry, regulators, and AI researchers.

\subsubsection{Holistic Evaluation for Industry Applications}

The open FinLLM leaderboard covers seven tasks: categories—Information Extraction (IE), Textual Analysis (TA), Question Answering (QA), Text Generation (TG), Risk Management (RM), Forecasting (FO), and Decision-Making (DM). Such a broad coverage allows industry professionals to identify models suited for specific applications, such as sentiment analysis (TA) for market predictions or risk management (RM) for credit scoring and fraud detection.

\subsubsection{Takeaway from Initial Evaluations}

Initial results show that general-purpose models like GPT-4 often underperform in financial-specific tasks such as Relation Extraction (IE) or Financial QA (QA). In contrast, models fine-tuned on financial datasets, such as FinLLaMA, perform better in tasks like stock movement forecasting (FO), suggesting that domain-specific training is critical for financial tasks. This highlights the need for organizations to focus on models tailored to financial data for high-stakes decision-making.

\subsubsection{Implications for Regulators}

Regulatory bodies, such as central banks or the SEC, could use the open FinLLM leaderboard's evaluations to assess AI models in financial institutions. The leaderboard provides a transparent way to evaluate models for tasks like fraud detection (RM) and compliance monitoring (IE), helping regulators establish baseline performance for AI models in finance. The zero-shot evaluations also ensure models are tested on unseen tasks, providing realistic assessments for evolving market conditions.

\subsubsection{AI Readiness in Finance}

From a research perspective, the open FinLLM leaderboard’s comprehensive evaluations highlight areas where financial LLMs excel and where further improvements are needed. The diverse financial tasks will encourage the development of AI models that are close to real-world adoption.

\section{Features}

This section describes the features of the open FinLLM leaderboard.

\subsection{Beyond Scores: Understanding Performance}

Our team’s extensive experience in the financial Large Language Model (LLM) sector has inspired us to take the next step in advancing this field: the development of the Open Financial LLM Leaderboard. While traditional benchmarks provide useful numerical scores for various tasks, they often fail to explain what these scores mean in practice, especially for users unfamiliar with financial-specific tasks. This realization motivated us to design a benchmark that not only delivers scores but also provides deeper insights into model performance and usability.

\subsubsection{From Experience to the Open Financial LLM Leaderboard}
\begin{figure}[t]
    \centering
    \includegraphics[width=0.5\textwidth]{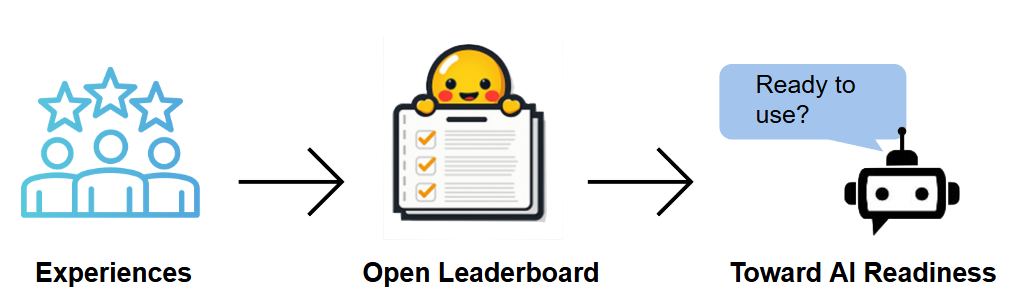}
    \caption{From user experiences to AI readiness: The iterative process of evaluating FinLLMs through the open leaderboard.}
    \label{fig:Experience2Readiness}
\end{figure}

Fig.~\ref{fig:Experience2Readiness} highlights our journey from leveraging user experiences to advancing AI readiness in financial applications. User feedback has been critical in shaping this project, allowing us to create a leaderboard that evaluates FinLLMs not only on their task-specific performance but also on their readiness for real-world financial tasks.

\subsubsection{Moving Beyond Scores}
\begin{figure}[t]
    \centering
    \includegraphics[width=0.5\textwidth]{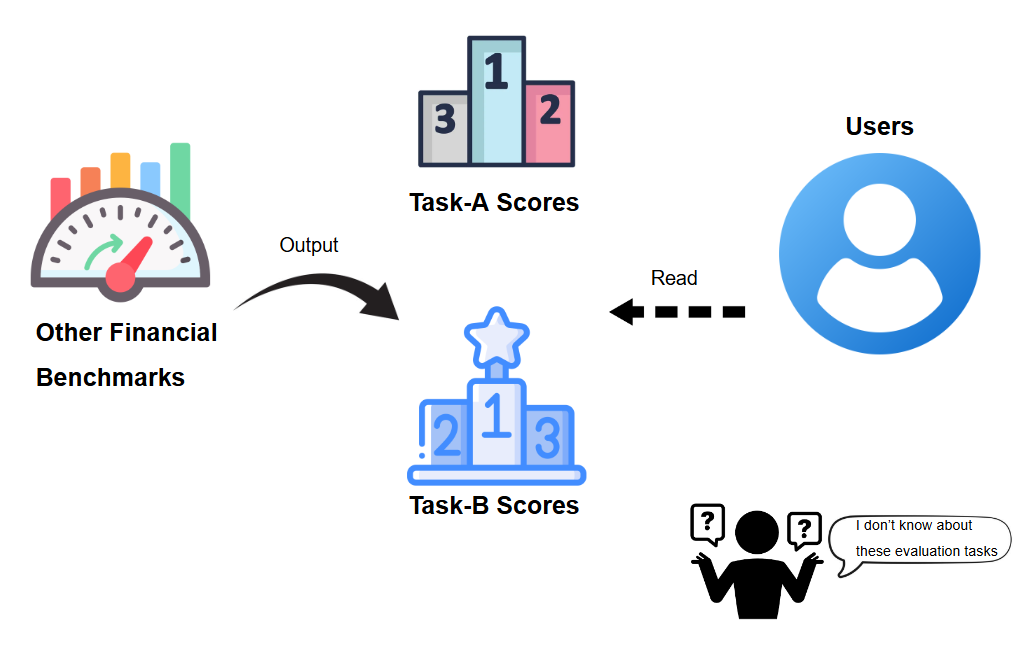}
    \caption{Integration of additional benchmarks: Traditional task-specific scores provide a foundation but may lack interpretability for non-expert users.}
    \label{fig:Other_Benchmark_diagram}
\end{figure}

Traditional benchmarks typically provide tables of task-specific scores, as shown in Fig.~\ref{fig:Other_Benchmark_diagram}. While these scores are useful for comparing models, they often lack context and practical interpretation for non-expert users. For example, users unfamiliar with financial tasks like sentiment analysis or compliance monitoring may struggle to understand how these scores relate to their specific needs.

To address this gap, we designed the Open Financial LLM Leaderboard to go beyond just delivering numerical scores. It provides insights into how models perform under specific financial scenarios, enabling users to connect scores to real-world applications.

This feature enables users to move beyond the surface-level scores and dive into qualitative aspects of LLM performance, such as accuracy, relevance, and susceptibility to errors like hallucinations or misinformation. For instance, users can identify whether a model correctly interprets financial terms or generates erroneous outputs in high-stakes scenarios.

\begin{figure}[t]
    \centering
    \includegraphics[width=0.5\textwidth]{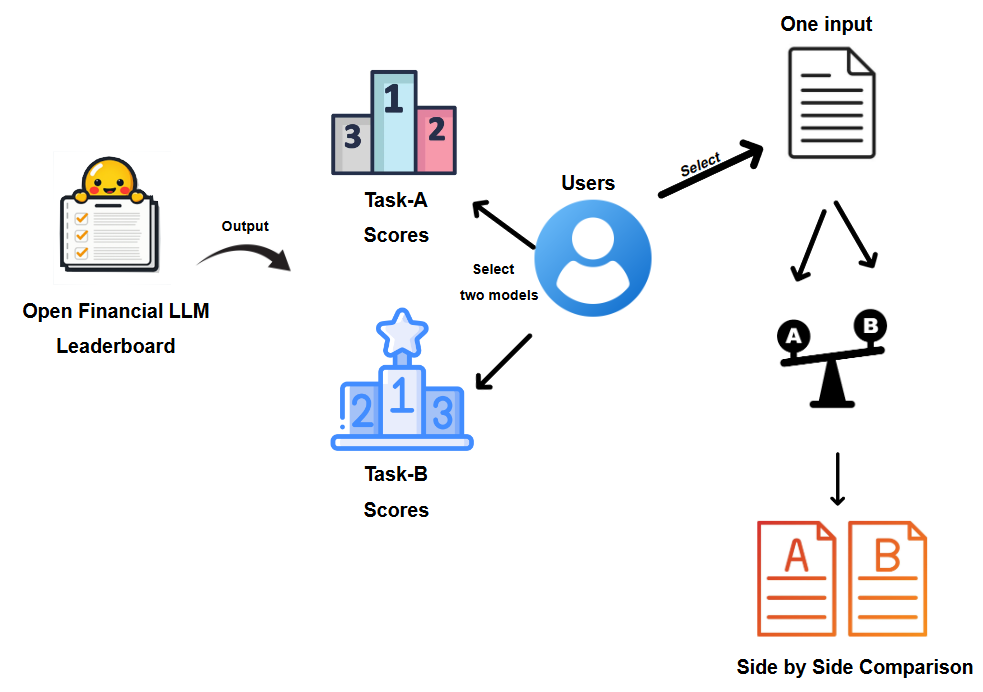}
    \caption{Evaluation workflow in the Open Financial LLM Leaderboard. Users can compare models side by side for deeper insights into their qualitative performance.}
    \label{fig:OFLLM_Benchmark_Diagram}
\end{figure}

\subsubsection{Opening the Black-Box of Financial Task Evaluation}

The inclusion of side-by-side comparisons helps us move beyond the black-box nature of traditional benchmarks. It allows for the identification of nuanced performance issues, such as:
\begin{itemize}[leftmargin=*]
    \item Error patterns related to hallucinations or misinformation.
    \item Strengths and weaknesses in interpreting financial documents.
    \item Practical usability for specific financial tasks like risk management or decision-making.
\end{itemize}

By integrating these features, the Open Financial LLM Leaderboard not only evaluates model accuracy but also enhances transparency, providing users with actionable insights for model selection and application.

\section{Use Scenarios}

\subsection{(Simple) Definitions and Financial Events}

\subsubsection{Definitions}
Financial concepts are not easy to understand. A chatbot could democratize financial knowledge to users with little background. 

\begin{figure}[t]
    \centering
    \includegraphics[width=0.53\textwidth]{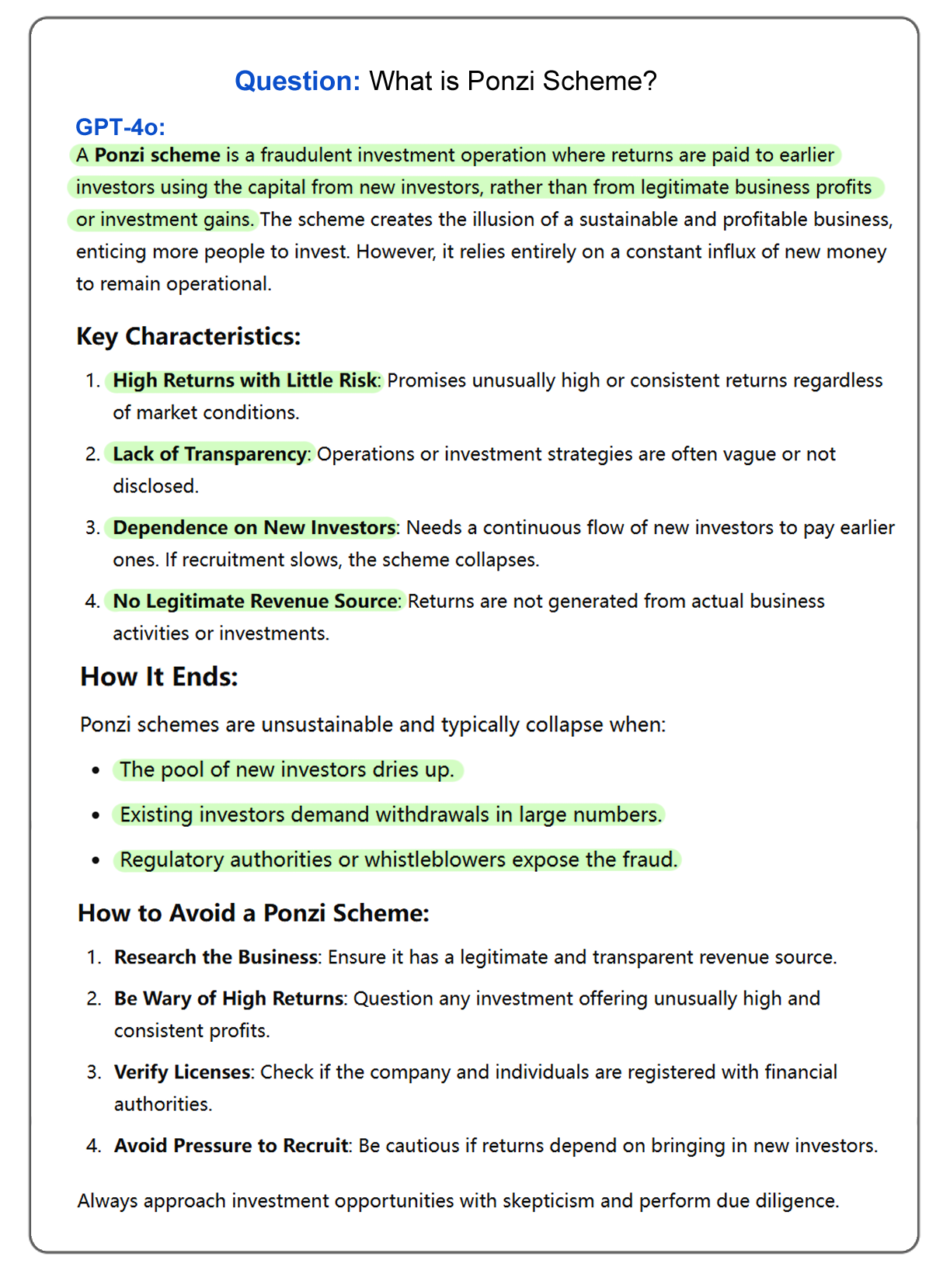}
    \caption{A simple Q\&A example where a user asks “What is a Ponzi Scheme?” Text highlighted in \textcolor{green}{Green} indicates correct information provided by GPT-4o.}
    \label{fig:ponzi_scheme_example}
\end{figure}

\begin{figure}[t]
    \centering
    \includegraphics[width=0.53\textwidth]{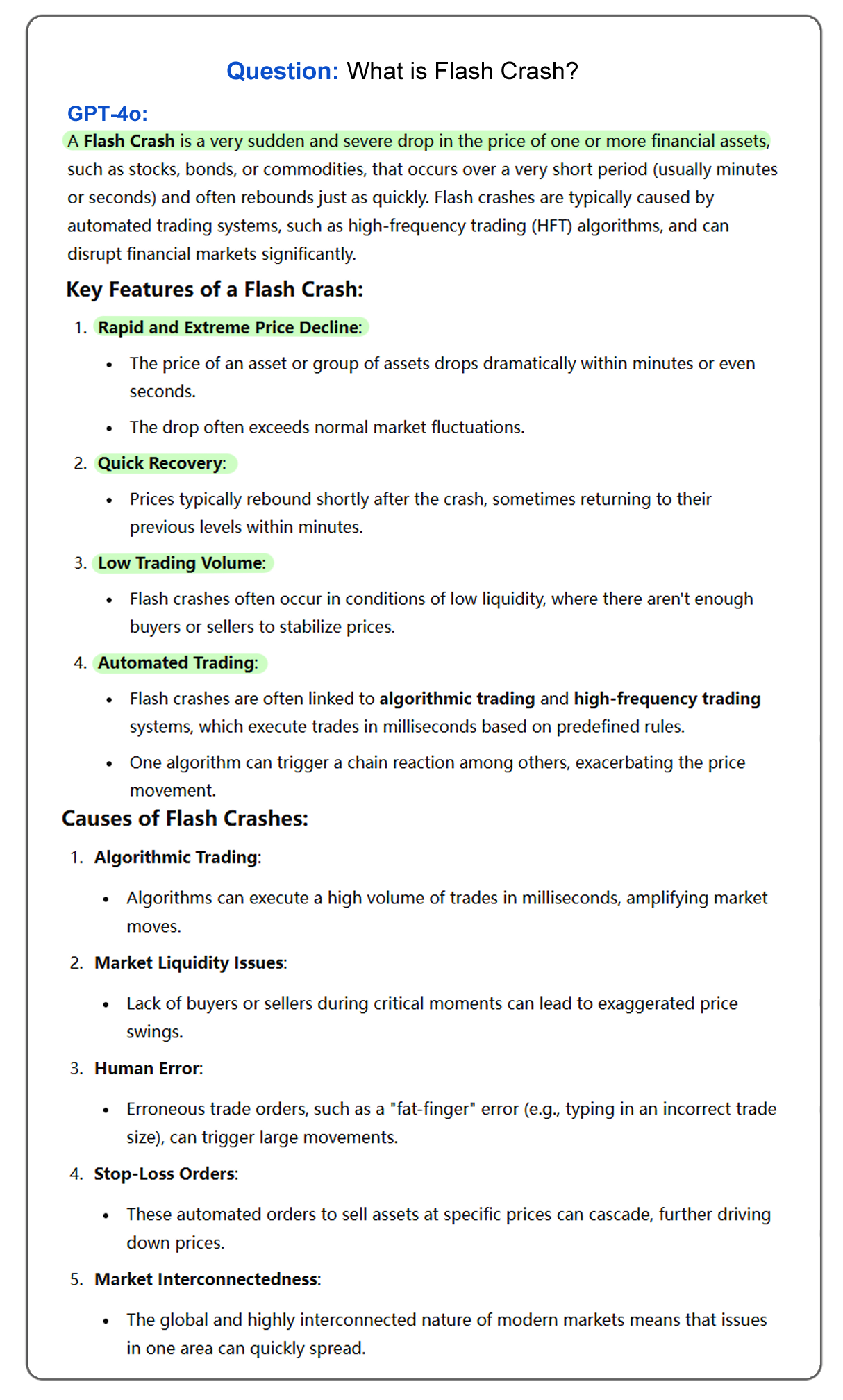}
    \caption{A Q\&A example about a Flash Crash. Text in \textcolor{green}{green} indicates correct explanatory details provided by the LLM.}
    \label{fig:flash_crash_example}
\end{figure}

LLMs can be a reliable source for general public to understand financial terms and definitions. Fig.~\ref{fig:ponzi_scheme_example} show how a chatbot explains: “What is a Ponzi Scheme?” The LLM responds with a definition and characteristics, where texts highlighted in green match the official definition of Ponzi scheme.  Fig.~\ref{fig:flash_crash_example} shows another example with “Flash Crash”. The LLM provides an explanation, where texts highlighted in green match widely accepted definitions.

\subsubsection{Understanding Financial Events}

\begin{figure}[t]
    \centering
    \includegraphics[width=0.53\textwidth]{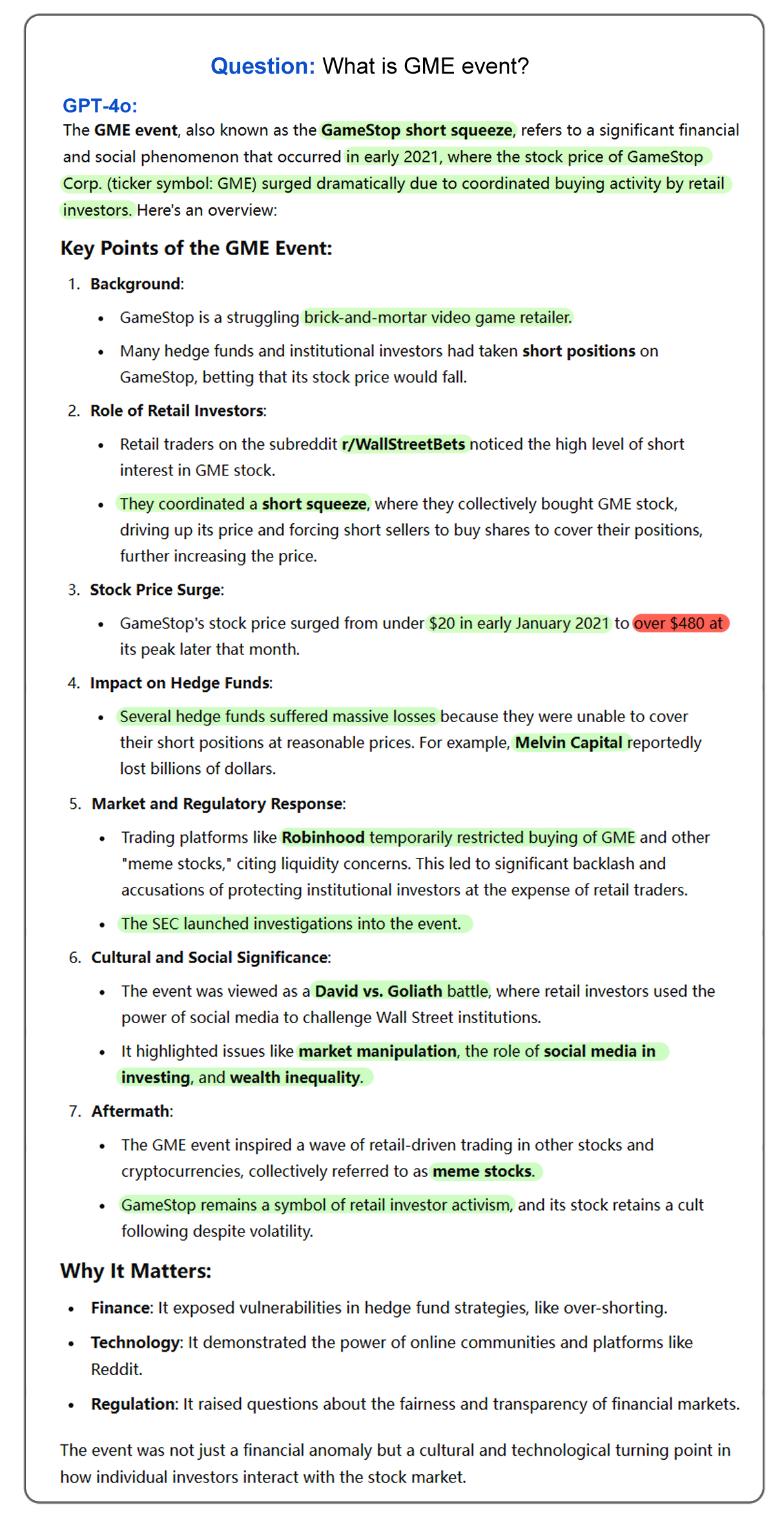}
    \caption{An example Q\&A about the GME event. Text in \textcolor{green}{Green} indicates correct conceptual explanations. Text in \textcolor{red}{Red} marks an incorrect numeric detail.}
    \label{fig:gme_event_example}
\end{figure}

LLMs can explain financial events and help public users understand the overall significance of the event. In Fig.~\ref{fig:gme_event_example}, the user asks about the GME event, and the LLM explains the event in ways matching the general understanding of the event’s cause and impact. Correct explanations are highlighted in green. However, when it comes to numbers, it inaccurately gives a peak price, shown in red. The actual peak price was around 95\$. While LLMs can accurately convey the core narrative of complex events, they could sometimes fail to give reliable numerical details.

\subsection{Efficiant Legal Consultation Prep}

We illustrate how using an LLM model to refine questions before a legal consultation, which facilitates more focused and efficient communication with lawyers, and thus helps users save time and reduce costs.

\subsubsection{Lawyer Consultation Cost Comparison}
\begin{figure}[t]
    \centering
    \includegraphics[width=0.5\textwidth]{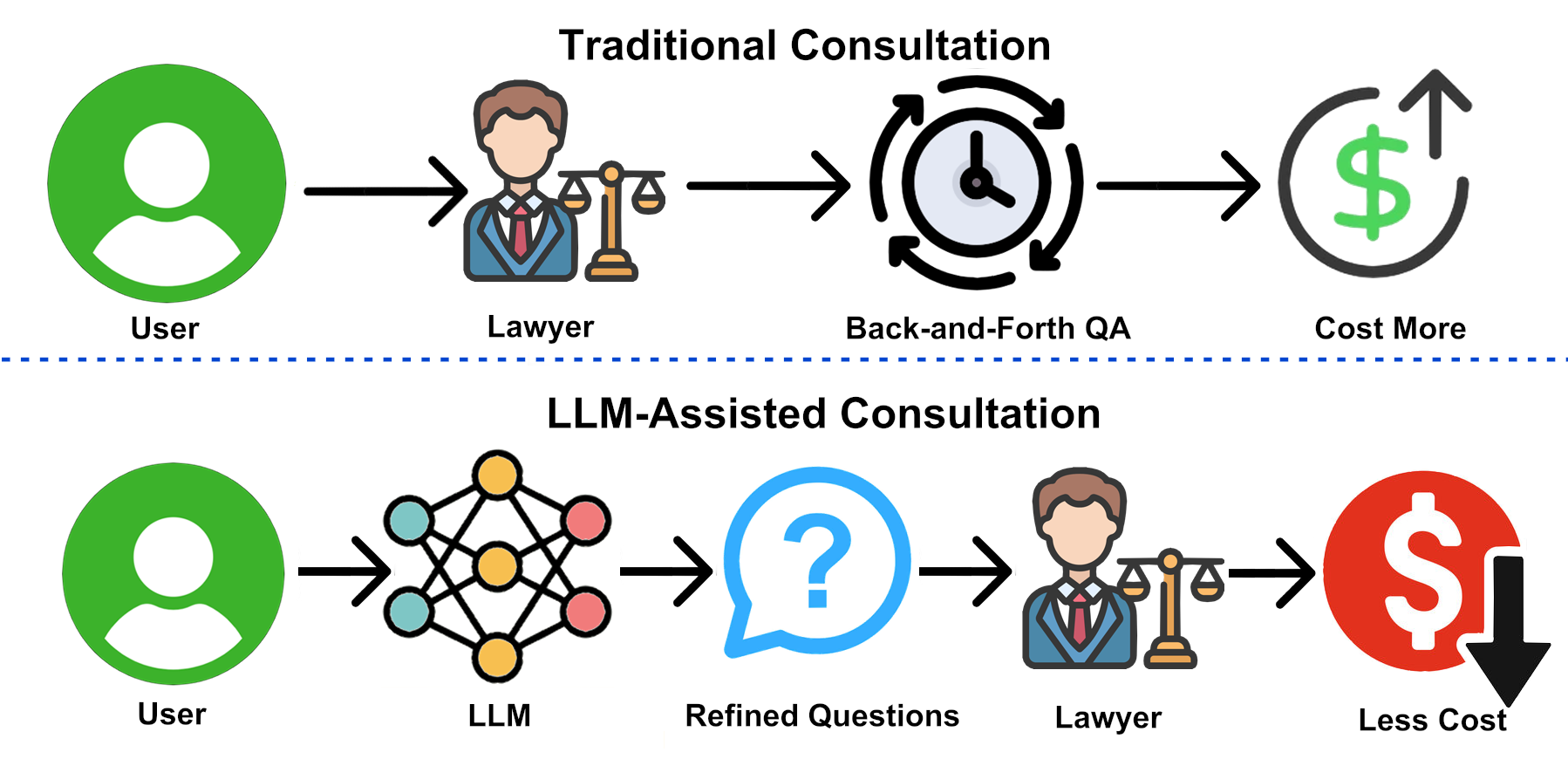}
    \caption{Comparison of lawyer consultation costs with and without pre-consultation question refinement using an LLM.}
    \label{fig:lawyer-use-scenario-diagram}
\end{figure}

Fig.~\ref{fig:lawyer-use-scenario-diagram} demonstrates the cost implications of using an LLM to refine questions before a legal consultation. In the traditional consultation workflow, the user engages in repetitive back-and-forth QA with the lawyer, which is time-consuming and leads to higher fees. By contrast, using an LLM to refine and organize questions beforehand enables more focused and efficient discussions with the lawyer, significantly reducing consultation time and associated costs. For example, without preparation, a consultation might last several hours, incurring substantial costs. However, with LLM assistance, the user can streamline the process, potentially reducing the session to a fraction of the time and lowering the overall cost.

\subsubsection{Example of Question Refinement}
\begin{figure*}[h]
    \centering
    \includegraphics[width=\textwidth]{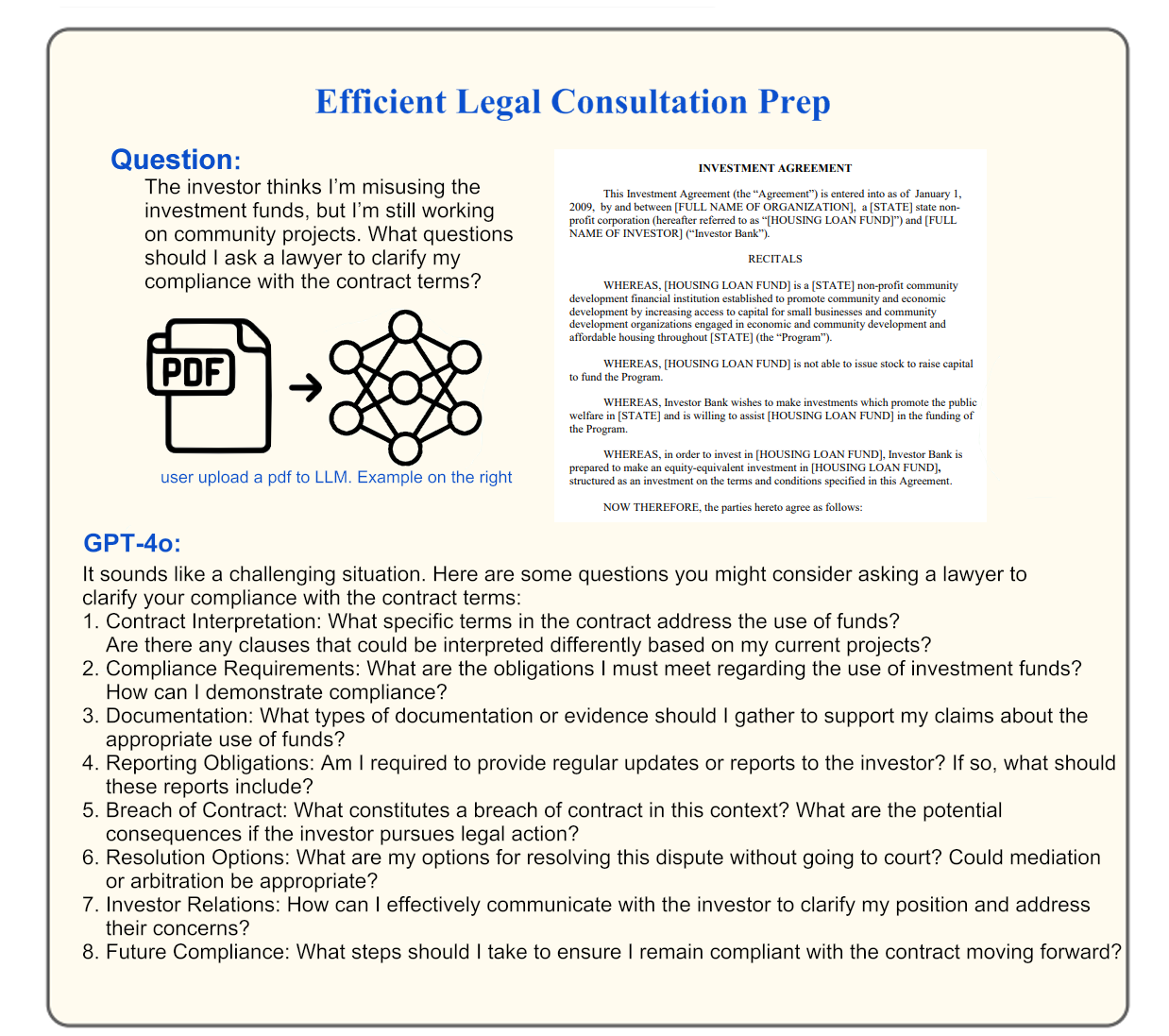}
    \caption{Prompt example and LLM-generated refined questions to aid in lawyer consultation regarding investment fund usage compliance.}
    \label{fig:lawyer-use-scenario-example}
\end{figure*}

Fig.~\ref{fig:lawyer-use-scenario-example} provides an example use scenario where the user suspects they might be non-compliant with investment fund usage terms and wants to clarify their position before consulting a lawyer. The LLM assists by generating a list of specific, refined questions to present to the lawyer, reducing the consultation time and ensuring that key contract compliance areas are covered.

\begin{figure*}[h]
    \centering
    \includegraphics[width=0.9\textwidth]{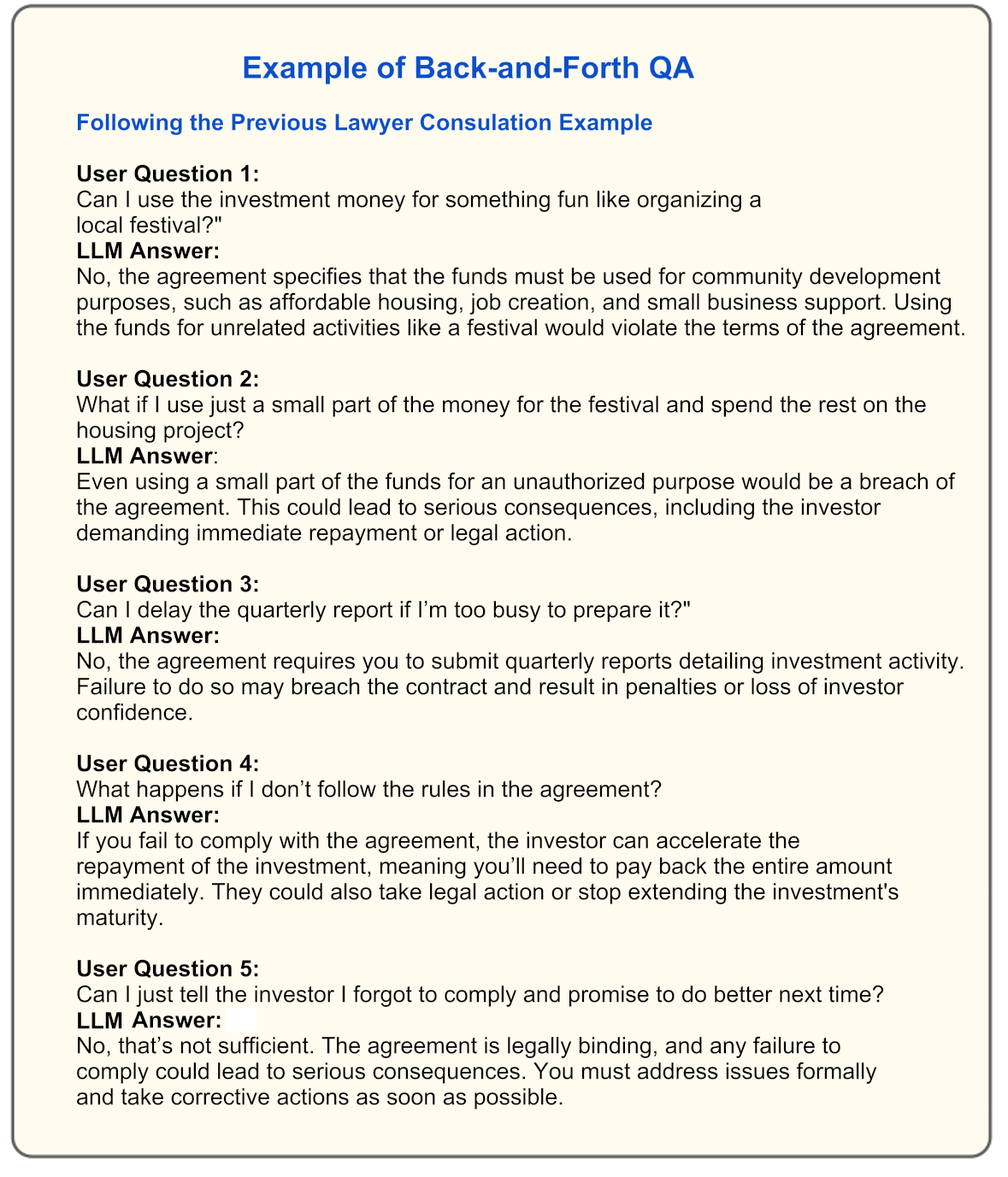}
    \caption{Example of inefficient back-and-forth QA. This scenario could occur between a user and a lawyer or an LLM. However, an LLM can handle these types of repetitive or basic questions without incurring additional costs. On the other hand, lawyers charge for their time, making such interactions expensive.}
    \label{fig:silly-user-BAF-QA-example}
\end{figure*}

\subsubsection{Detailed Analysis of Back-and-Forth QA}

Fig.~\ref{fig:silly-user-BAF-QA-example} illustrates inefficiencies in back-and-forth QA interactions, especially when users lack financial or legal expertise. Below is a concise analysis of the example.

\paragraph{Key Inefficiencies:}
\begin{itemize}
    \item \textbf{Repetitive Questions:} The user repeatedly asks similar questions (e.g., misusing funds for a festival),.
    \item \textbf{Basic Concept Clarifications:} Queries about fundamental terms (e.g., reporting obligations) show a lack of preparation.
    \item \textbf{Unrealistic Proposals:} Suggestions like simply promising compliance reflect unrealistic expectations.
    \item \textbf{Cost Implications:} Lawyers charge for their time, making inefficient interactions expensive.
\end{itemize}

\paragraph{Benefits of Using an LLM:}
\begin{itemize}
    \item \textbf{Cost Savings:} LLMs handle repetitive or basic queries at no cost.
    \item \textbf{Preparation:} LLMs help refine user questions, streamlining subsequent lawyer consultations.
    \item \textbf{Availability:} Instant, 24/7 responses reduce delays.
\end{itemize}

\subsection{General Publics vs. LLM-Assisted Financial Document Analysis}

\begin{figure}[t]
    \centering
    \includegraphics[width=0.5\textwidth]{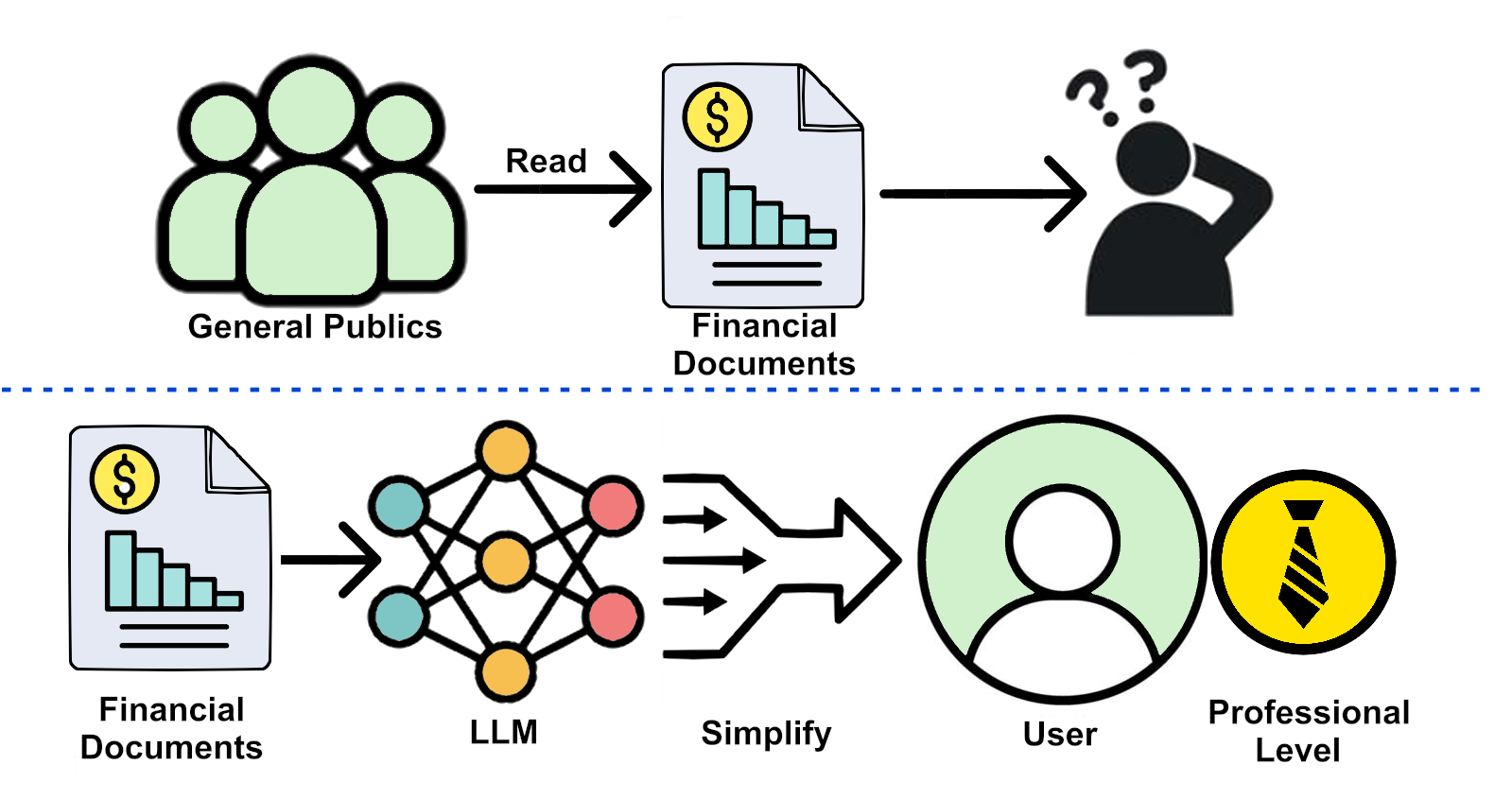}
    \caption{General public vs. LLM-assisted financial document analysis. FinLLMs will upgrade the financial analytical capability of the general public to a professional-level.}
    \label{fig:report_use_scenario_diagram}
\end{figure}

Fig.~\ref{fig:report_use_scenario_diagram} illustrates the difference between how the general public and an LLM-assisted user approach financial documents. Without assistance, general users may struggle to understand complex financial reports, leading to confusion and limited comprehension. However, with the help of an LLM, financial documents are simplified, enabling the general public to achieve a professional-level understanding and make more informed decisions.

\subsubsection{Example of Financial Report Analysis}

\begin{figure*}[t]
    \centering
    \includegraphics[width=\textwidth]{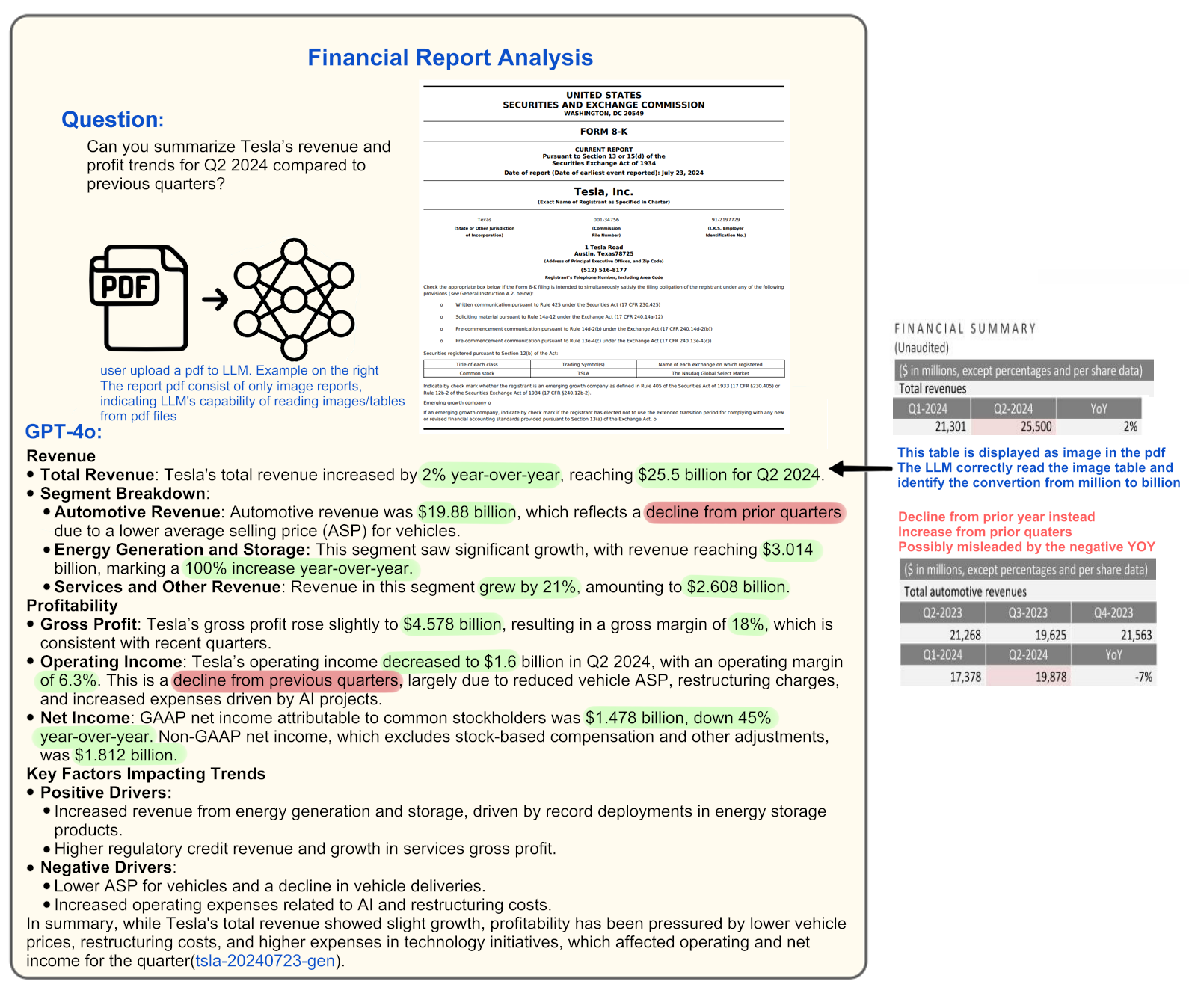}
    \caption{Example of financial report analysis with highlighted errors and corrections. Text marked in *red* indicates inaccuracies in the LLM's interpretation, such as misunderstanding revenue trends or being misled by year-over-year (YOY) changes. Text marked in *green* showcases accurate information, reflecting the LLM's correct analysis of numerical data and trend direction.}
    \label{fig:report_use_scenario_example}
\end{figure*}

Fig.~\ref{fig:report_use_scenario_example} provides an example scenario where the user uploads a financial report and prompts the LLM for a summary of Tesla’s revenue and profit trends for Q2 2024. The LLM generates a summary that highlights revenue breakdown, profitability, and key factors affecting trends, helping the user to quickly understand complex financial document.

\subsection{AI-Driven Regulatory Compliance in Capital Markets} \label{subsec:trade_data_quality}

\subsubsection{Data Quality Challenges in Capital Markets}
Within the Financial Services industry, and especially within capital markets, ensuring data quality remains a persistent challenge \textit{even with} standardized data models and agreed formats for communicating derivative lifecycle details. Key issues include:
\begin{itemize}
    \item Heterogeneous handling of trade attributes/fields across organizations
    \item Regulatory mandates for trade repositories to normalize disparate data
    \item System-to-system variations in data field implementations
\end{itemize}

\subsubsection{End-to-End Solution Architecture}
We demonstrate an AI-driven pipeline linking unstructured regulatory texts to actionable data quality insights through four key stages:

\begin{figure}[htbp]
    \centering
    \includegraphics[width=0.5\textwidth]{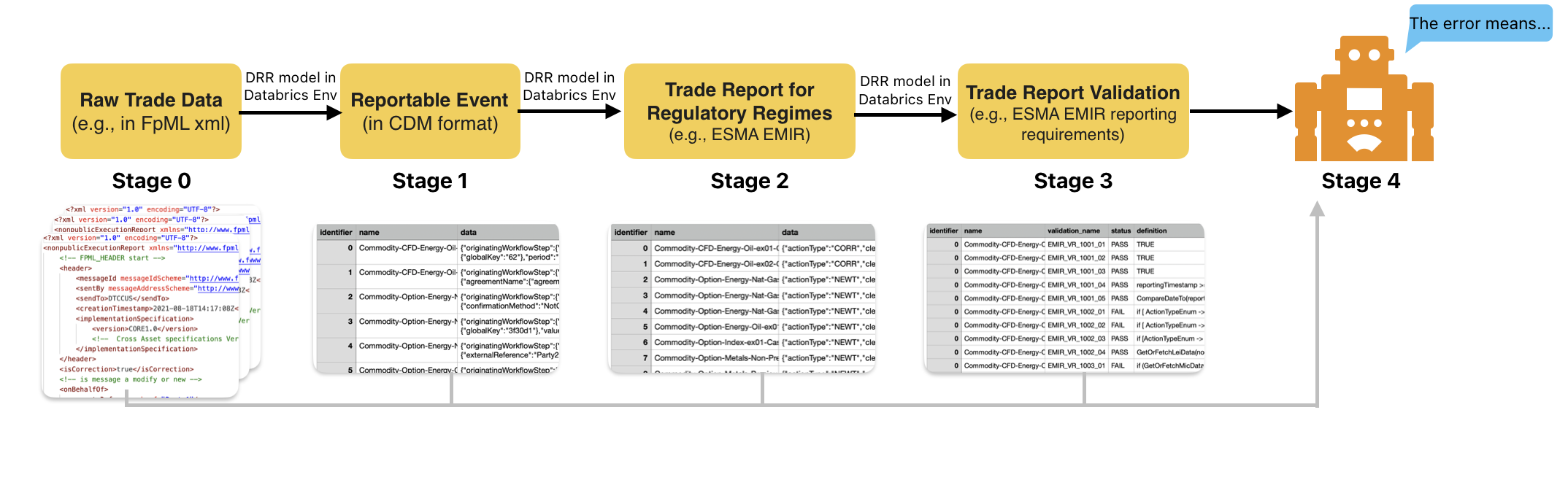}
    \caption{Data quality resolution workflow showing Stage 0 (Raw FpML data) through Stage 4 (Chatbot-assisted error correction).}
    \label{fig:error_resolution_diagram}
\end{figure}

\begin{enumerate}
    \item \textbf{Stage 0: Raw Trade Data} - FpML XML from counterparty systems
    \item \textbf{Stage 1: CDM Conversion} - JSON transformation using Digital Regulatory Reporting (DRR) engine
    \item \textbf{Stage 2: Regulatory Mapping} - Generation of ESMA EMIR reports
    \item \textbf{Stage 3: Rule Validation} - Automated checks against EMIR validation rules
    \item \textbf{Stage 4: Error Resolution} - Chatbot-assisted diagnosis and correction
\end{enumerate}

\begin{figure*}[htbp]
    \centering
    \includegraphics[width=0.8\textwidth]{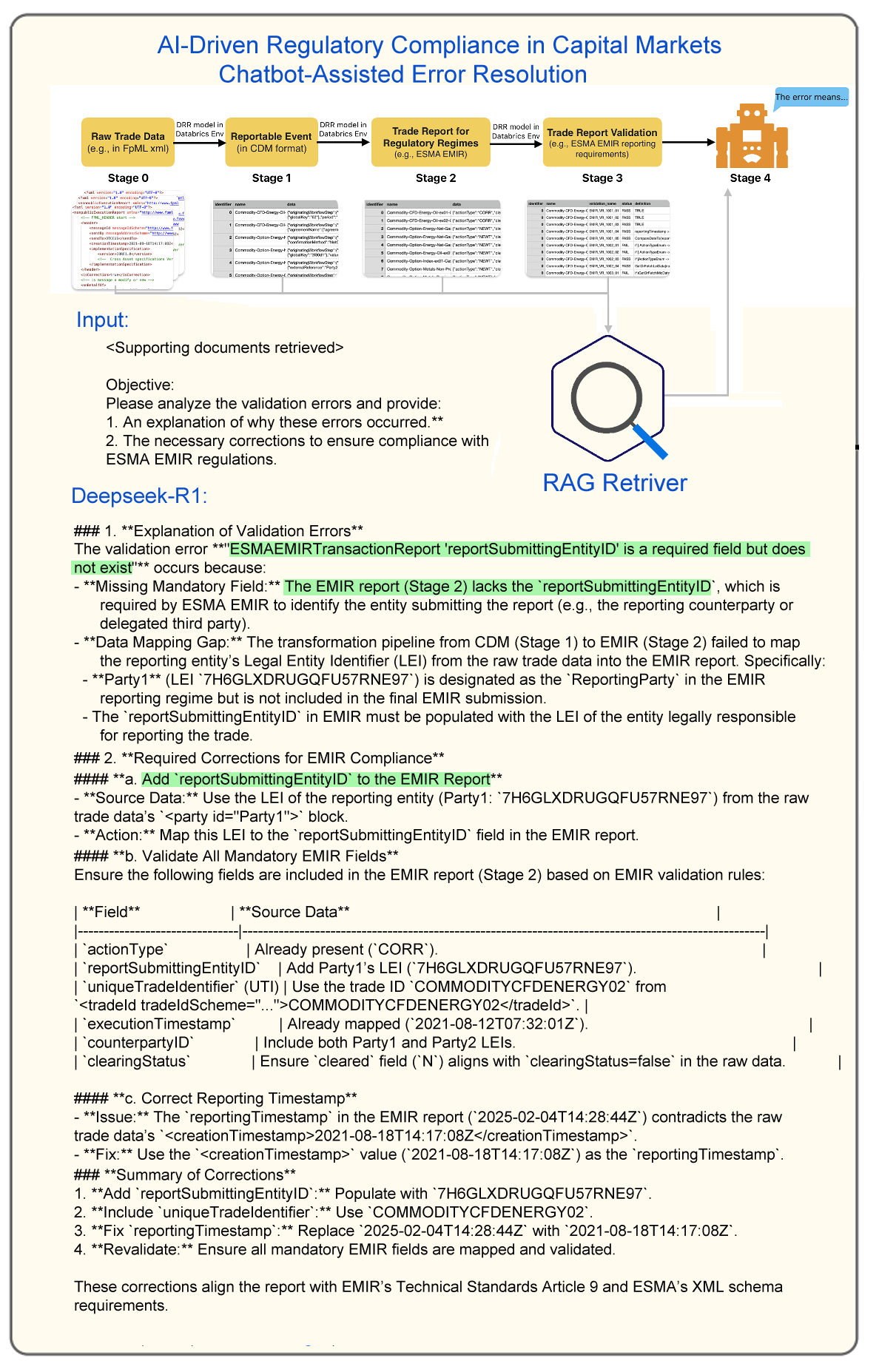}
    \caption{Example of error resolution with highlighted errors and corrections. Text marked in \textcolor{green}{Green} showcases correct information and corrections identified by LLM.}
    \label{fig:error_resolution_example}
\end{figure*}

\subsubsection{Implementation Example}
Figure~\ref{fig:error_resolution_example} demonstrates the error resolution process for a missing \texttt{reportSubmittingEntityID} field, a common EMIR compliance failure:

\subsection{XBRL Analytics} \label{app:failure_cases_xbrl_analytics}

A common scenario for XBRL analysis is analyzing financial statements to calculate financial ratios or to make financial forecasts.

\subsubsection{\textbf{Failure Case: ROA Calculation}}
\begin{figure*}[t]
    \centering
    \includegraphics[width=\textwidth]{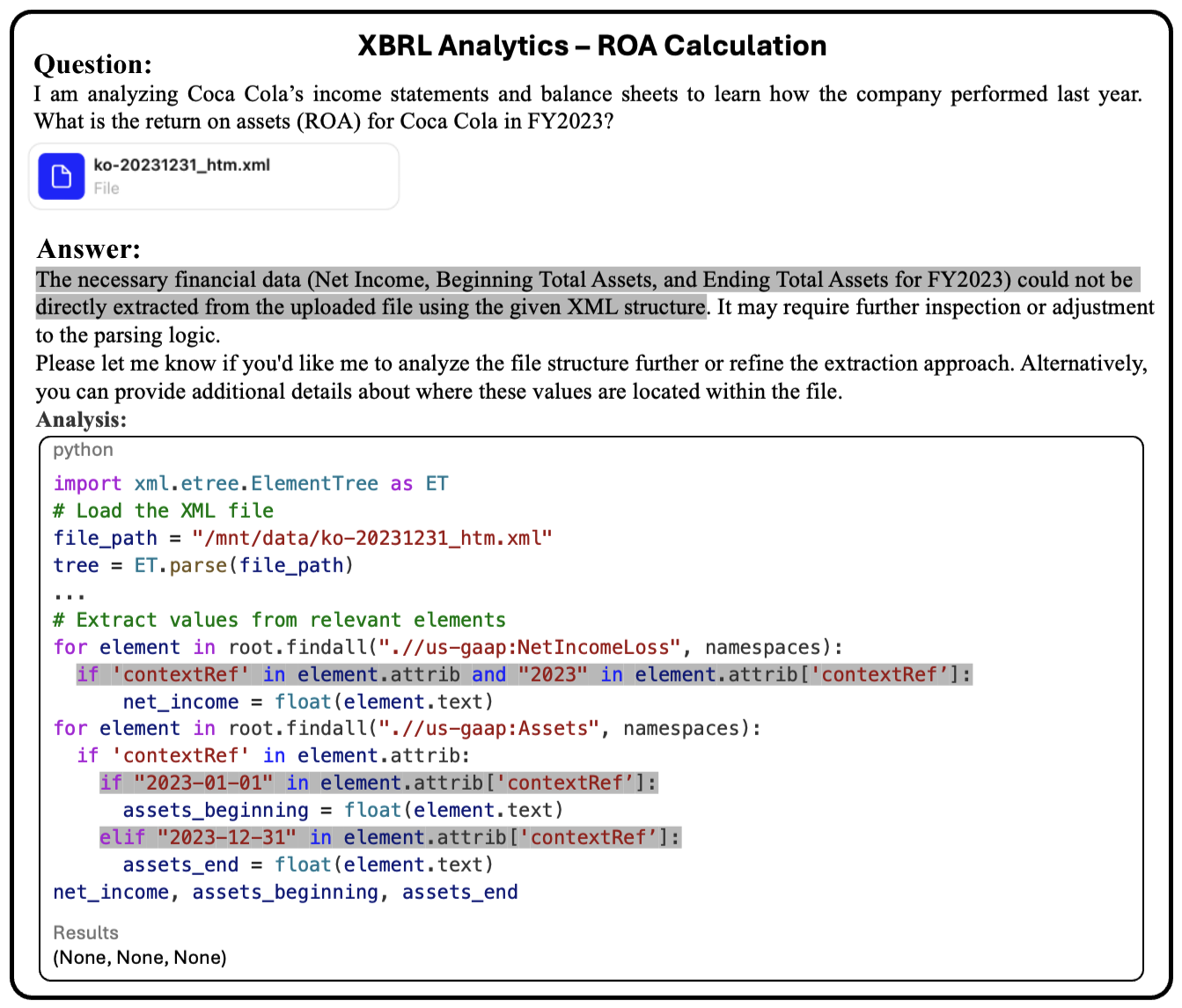}
    \caption{A failure case for XBRL analytics. The model is GPT-4o. The XBRL file is uploaded. Sentences in gray are where failure occurs. The model cannot parse the XBRL file correctly. In its code for analysis, it doesn't use \texttt{contextRef} correctly, which is not 'yyyy-mm-dd' or 'yyyy' format.}
    \label{fig:failure_xbrl_1_full_1}
\end{figure*}

\begin{figure*}[t]
    \centering
    \includegraphics[width=\textwidth]{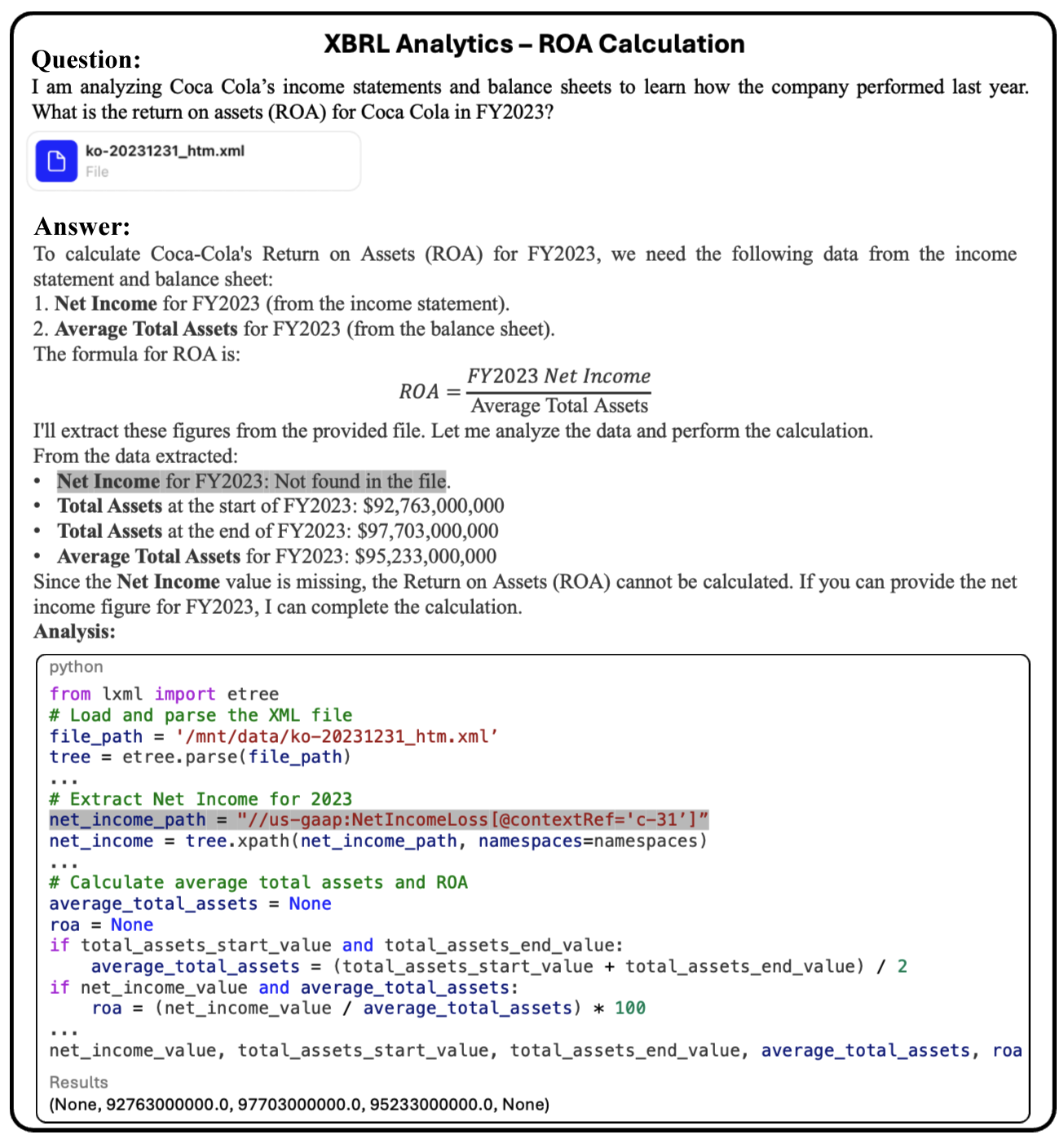}
    \caption{A failure case for XBRL analytics. The model is GPT-4o. The XBRL file is uploaded. Sentences in gray are where failure occurs. The model cannot identify the contexts for net income correctly. In its code for analysis, it uses the incorrect \texttt{contextRef} of "c-31", which should be "c-1"}
    \label{fig:failure_xbrl_1_full_2}
\end{figure*}

\begin{figure*}[t]
    \centering
    \includegraphics[width=\textwidth]{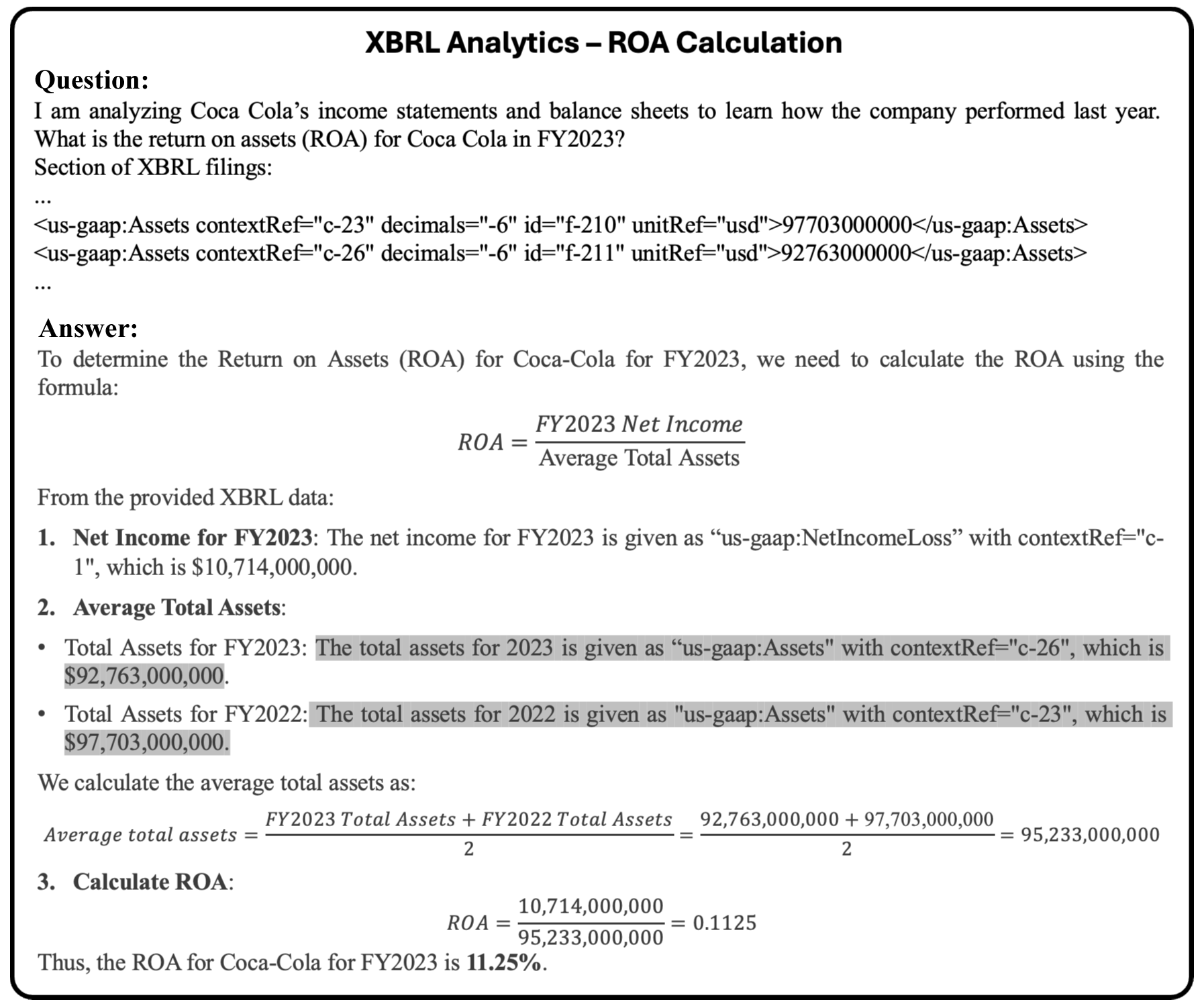}
    \caption{A failure case for XBRL analytics. The model is GPT-4o. Not all XBRL content is shown in the picture. Sentences in gray are where failure occurs. The \texttt{contextRef} and values for total assets are recognized wrongly. Fig. \ref{fig:failure_xbrl_statements} shows the income statements in the presentation mode. The \texttt{contextRef} should be “c-23” and the value is \$97,703,000,000 for FY2023. The \texttt{contextRef} should be “c-26” and the value is \$92,763,000,000 for FY2022.}
    \label{fig:failure_xbrl_1_section}
\end{figure*}

Figs. \ref{fig:failure_xbrl_1_full_1}, \ref{fig:failure_xbrl_1_full_2} and \ref{fig:failure_xbrl_1_section} show some failure cases when calculating Coca-Cola's
return on assets (ROA) in FY2023. Fig. \ref{fig:failure_xbrl_statements} \footnote{Obtained from 10-K filings from \url{https://www.sec.gov/edgar/browse/?CIK=21344&owner=exclude}} (a) and (b) shows the parts in financial statements related to this analysis. The model used is GPT-4o.

\textbf{XBRL File Parsing}. As shown in Fig. \ref{fig:failure_xbrl_1_full_1}, we send the query of Coca-Cola's ROA in FY2023 and upload the XBRL instance file (in XML format) \footnote{\url{https://www.sec.gov/Archives/edgar/data/21344/000002134424000009/ko-20231231_htm.xml}}. The model fails to parse the XBRL file and extract corresponding values. In the model's generated code, it uses the incorrect "yyyy-mm-dd" or "yyyy" format for \texttt{contextRef} to extract values. In this XBRL file, \texttt{contextRef} is in the "c-n" format where $n$ is the assigned sequence number.

\textbf{Contexts}. There are failure cases in identifying the contexts:
\begin{itemize}[leftmargin=*]
    \item \textbf{Query with full information}. With the same query and file as above, as shown in Fig. \ref{fig:failure_xbrl_1_full_2}, although the model parses the XBRL file successfully, it fails to identify the context reference of net income in FY2023. In its generated code to extract the net income, the model uses the incorrect \texttt{contextRef} of "c-31," where the correct reference is "c-1." This mistake may be due to the noisy information in the file: "c-31" is a specific context reference used in the statement of shareholders' equity in FY2023, while "c-1" is the general context reference for FY2023. 
    \item \textbf{Query with specific information}. To provide more specific information, as shown in Fig. \ref{fig:failure_xbrl_1_section}, we concatenate the query with the income statement and balance sheet in FY2023 from the XBRL instance file in XML format. Although the model gives the correct final result, it makes a mistake in recognizing the context reference and values for total assets in FY2022 and FY2023. The correct \texttt{contextRef} and value for FY2023 is "c-23" and $\$97,703,000,000$, and "c-26" and $\$92,763,000,000$ for FY2022.
\end{itemize}

\textbf{Concepts}. We also ask the related concepts in follow-up questions, including the definition of FY2023, the meaning of "KO" in the name of the uploaded file, the tag "us-gaap:NetIncomeLoss" and "us-gaap:Assets," and the unit measure of net income and total assets. The model's responses are correct.

\textbf{Financial Ratio Formula}. We also ask about the source of the formula for ROA. The model derives the ROA formula from its internal knowledge base, which is built on principles of financial analysis and accounting learned during training. 

\textbf{Calculation}. In the case when the full file is uploaded, as shown in Fig. \ref{fig:failure_xbrl_1_full_2}, the model generates and executes code to perform the calculation. When the query concatenated with the file sections is sent as input, the model performed the calculation itself based on the data provided in the XBRL tags and the financial formula.


\textbf{Other settings}.
Here we show the analysis for Coca Cola’s ROA in FY2023 in different settings.
\begin{itemize}[leftmargin = *]
    \item \textbf{Closed-book testing}. We first test GPT-4o in a closed-book setting, without financial statements or financial ratio formula provided in the query and without online search function. When directly asked about the ROA of Coca Cola in FY2023, GPT-4o fails to give the answer, saying that it has no access to real-time financial data for Coca-Cola. It only gives instructions to find the financial statements in the annual reports and gives a formula to calculate ROA, the net income divided by the total average assets for the period. 
    \item \textbf{Closed-book testing of stock ticker}. We replace the firm name "Coca Cola" with its stock ticker "KO" in the query. The model can successfully recognize the company as The Coca-Cola Company (KO), but still fails to give the answer with the same reason.
    \item \textbf{Online search}. We send the same query to GPT-4o with online search function. The model returns the values of financial items, the formula, and the ROA of 10.97\%, which are directly obtained from the Stock Data Online website \footnote{Retrieved on Nov 13, 2024 from \url{https://stock-data.online/stock/ko/profitability-ratio/return-on-assets}}, without performing calculations. 
    The formula for ROA on the website is to divide the net income in FY2023 (\$10,714,000) by the assets at the end of FY2023 (\$97,703,000). A more accurate formula for ROA is to divide the net income by the average total assets at the end and beginning of the year. Although all the values needed for the more accurate calculation are available on the website and the model correctly answers the formula in the closed book setting, GPT-4o still gets the inaccurate results directly from the website.

    \item \textbf{Online search with formula}. We send the query and formula to GPT-4o with online search function. The model obtains the value of financial terms from Macrotrends webiste \footnote{Retrived on Nov 13, 2024 from \url{https://www.macrotrends.net/stocks/charts/KO/cocacola/total-assets}}. Then it performs the calculation itself based on the provided formula. The returned result is $11.25\%$, which is correct.
    \item \textbf{Chain-of-thought (CoT)}. When only the query is sent to ChatGPT-o1-preview, the model follows the steps of assessing knowledge, ensuring data accuracy, assessing current financial status, assessing ROA, and assessing current data. It doesn't have knowledge about FY2023 and provides the ROA in FY2022. However, its data for FY2022 are inaccurate. We then send the query concatenated with the related segments in the XBRL file \footnote{ChatGPT-o1-preview currently doesn't support file uploading and online search}, and the model performs the analytics correctly. 
\end{itemize}

\subsubsection{\textbf{Failure Case: Revenue Forecasting}}

\begin{figure*}[t]
    \centering
    \includegraphics[width=\textwidth]{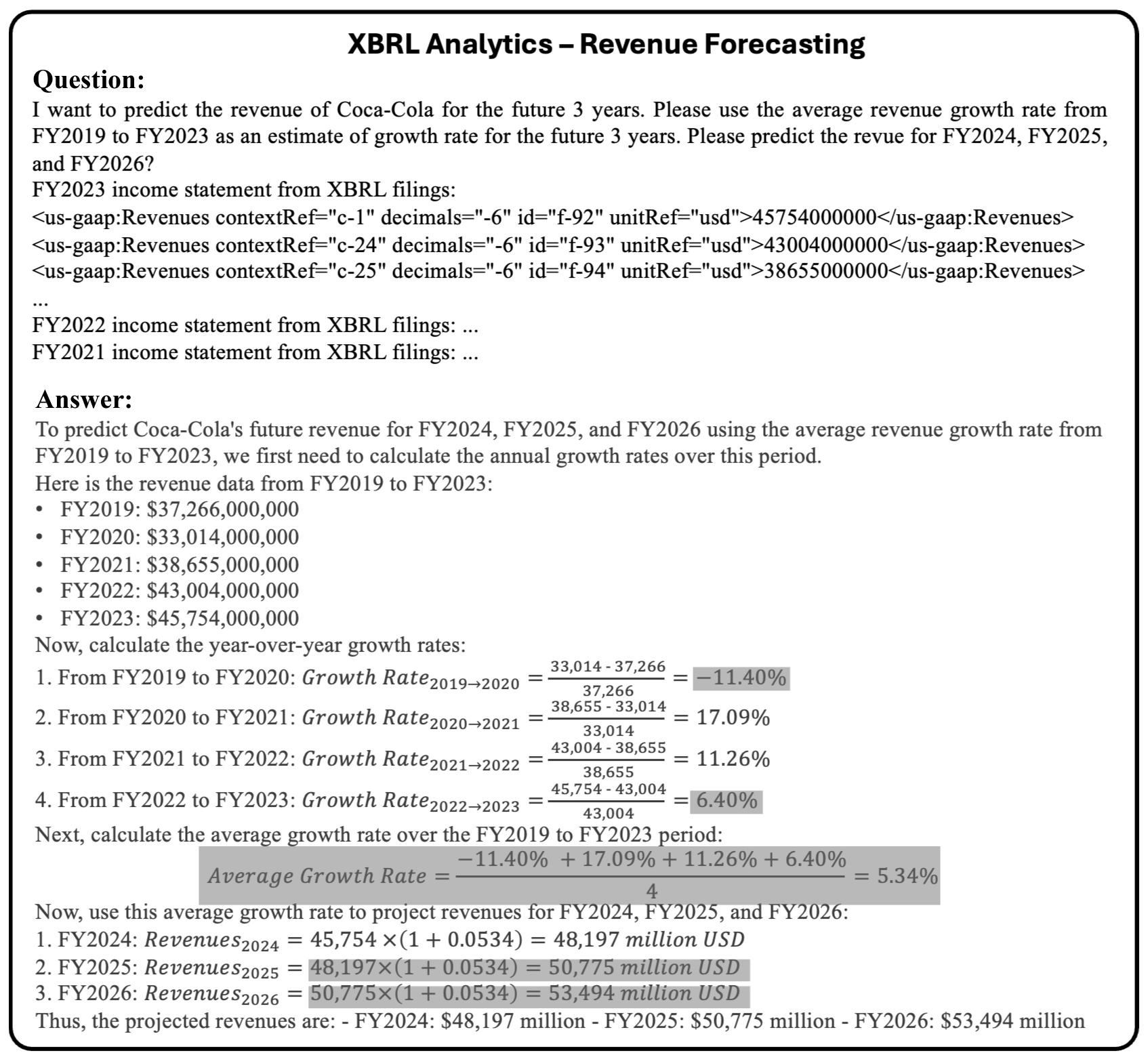}
    \caption{A failure case of financial math calculations in XBRL filings. The model is GPT-4o and it makes mistakes when calculating the formula in gray. Fig. \ref{fig:failure_xbrl_statements} shows the income statements.}
    \label{fig:failure_xbrl_3}
\end{figure*}

Fig. \ref{fig:failure_xbrl_3} shows a failure case of revenue forecasting for Coca-Cola. The task is to use the average revenue growth rate from FY2019 to FY2023 to predict revenues in the future 3 years. The income statements for FY2021, FY2022, and FY2023 from XBRL instance files are concatenated to the query. The model used is GPT-4o.

\begin{figure*}[t]
    \centering
    \includegraphics[width=\textwidth]{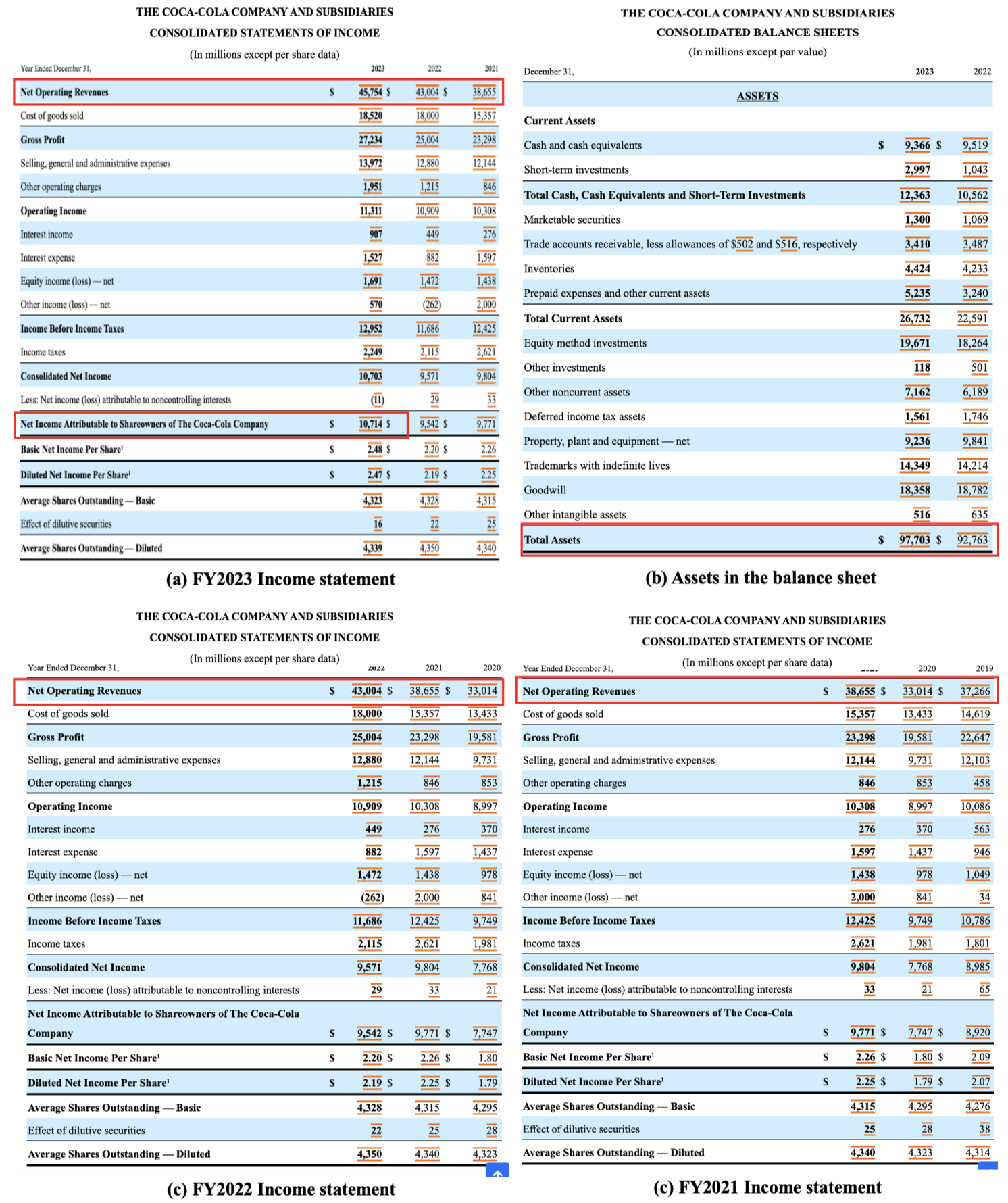}
    \caption{Parts from the financial statements of Coca-Cola.}
    \label{fig:failure_xbrl_statements}
\end{figure*}

\textbf{Context}. The model correctly identifies and extracts revenues from FY2019 to FY2023. Fig. \ref{fig:failure_xbrl_statements} (a), (c), and (d) shows the Coca-Cola's income statements.

\textbf{Financial formula}. The model uses the correct formulas to calculate the growth rate, the average growth rate, and the predicted revenue. 

\textbf{Calculation}. The model makes some mistakes in its calculation process, as highlighted in gray in Fig. \ref{fig:failure_xbrl_3}. When calculating the growth rates of FY2020 and FY2023, the model doesn't round the answers correctly. The correct roundings are $-11.41\%$ and $6.39\%$, respectively. In the calculation of the average growth rate, using values obtained from the previous steps, the answer $5.34\%$ is wrong, which should be $5.84\%$. Furthermore, when calculating revenue for FY2025 and FY2026 using the values in the formula, the answers are wrong, where $48,197 \times (1+0.0534) \approx 50,770$ and  $50,775 \times (1+0.0534) \approx 53,486$. If all calculations are correct, the predicted revenue is approximately $\$48,422$ million for FY2024, $\$51,245$ million for FY2025, and $\$54,232$ million for FY2026.

\begin{table*}[htbp]
\centering
\includegraphics[width=0.8\textwidth]{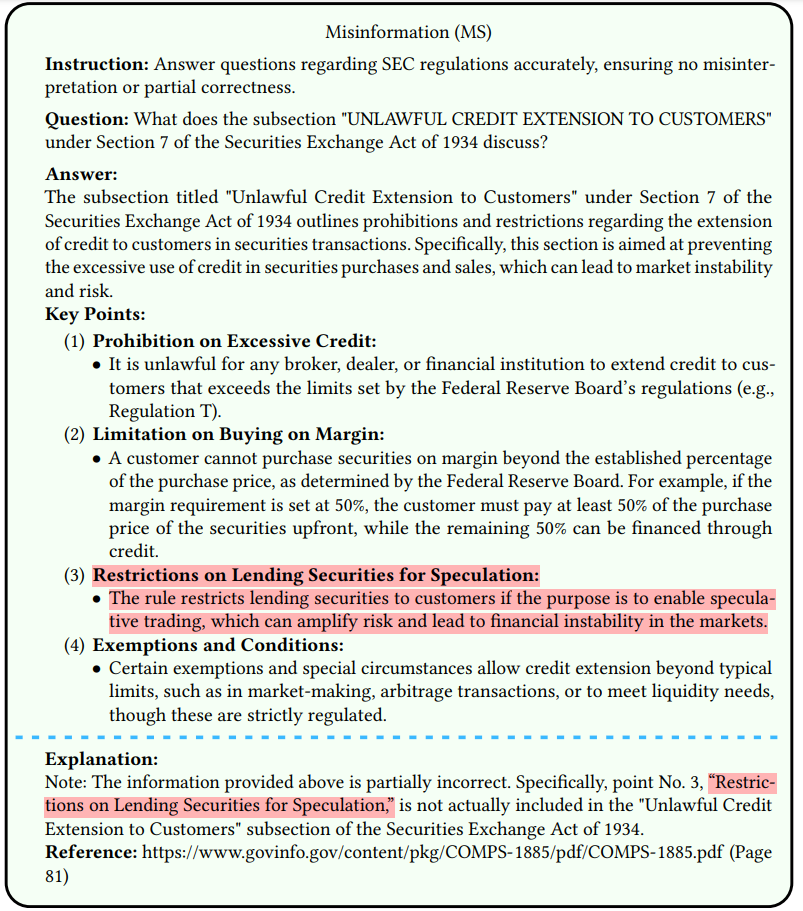}
\caption{An example of the Misinformation (MS) task, regarding misinformation correction in the SEC regulation context. Texts in \textcolor{red}{red} indicate the non-existing infomation generated by the LLM.}
\end{table*}


\begin{table*}[htbp]
\centering
\includegraphics[width=0.8\textwidth]{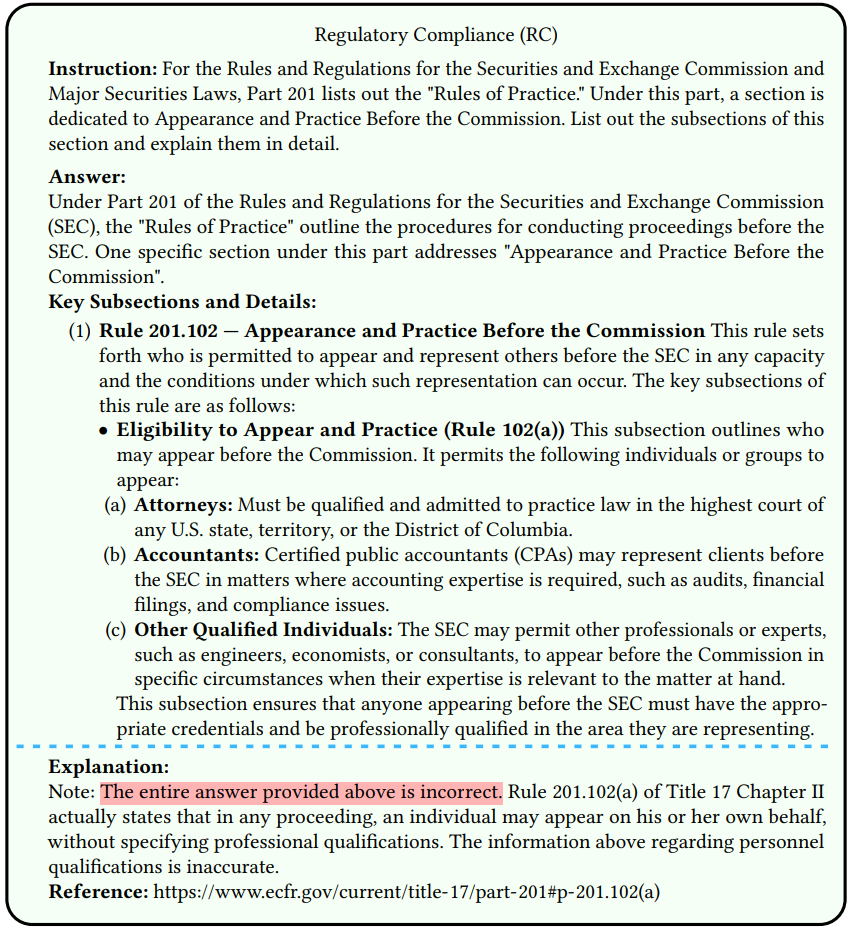}
\caption{An example of the Regulatory Compliance (RC) task, regarding SEC Rules of Practice related to ``Appearance and Practice before the Commission".}
\end{table*}

\end{document}